\documentclass[pdflatex,sn-mathphys-num]{arxiv-sn-jnl}
\usepackage[T1]{fontenc}
\usepackage{lmodern}
\usepackage{amsmath,amssymb,booktabs,graphicx,array,tabularx,longtable}
\usepackage{geometry}
\usepackage{xurl}
\makeatletter
\renewcommand\normalsize{\@setfontsize\normalsize{11bp}{16.5bp}\abovedisplayskip 10pt plus2pt minus2pt\belowdisplayskip 10pt plus2pt minus2pt\abovedisplayshortskip 4pt\belowdisplayshortskip 5pt\let\@listi\@listI}
\renewcommand\small{\@setfontsize\small{10bp}{13bp}}
\renewcommand\footnotesize{\@setfontsize\footnotesize{9bp}{11.5bp}}
\def\abstractheadfont{\reset@font\fontsize{11bp}{13.5bp}\bfseries\selectfont\titraggedcenter}
\def\abstractfont{\reset@font\fontsize{11bp}{13.5bp}\selectfont\leftskip=0pt\rightskip=0pt\parfillskip=0pt plus 1fil}
\def\keywordfont{\reset@font\fontsize{10bp}{12bp}\selectfont\leftskip=0pt\rightskip=0pt plus0.5fill}
\def\bibfont{\reset@font\fontfamily{\rmdefault}\fontsize{11bp}{13.5bp}\selectfont}
\makeatother
\providecommand{\burl}[1]{\url{#1}}

\normalsize
\renewcommand{\arraystretch}{1.16}

\hypersetup{colorlinks=true,linkcolor=black,citecolor=black,urlcolor=blue,bookmarksdepth=2,pdftitle={Model Discovery for Virtual Chaos Engineering with Incomplete Tracing: An eBPF Study},pdfauthor={Anatoly A. Krasnovsky}}

\begin{document}

\title{Model Discovery for Virtual Chaos Engineering with Incomplete Tracing: An eBPF Study}
\author[1,2]{\fnm{Anatoly A.} \sur{Krasnovsky}}
\affil[1]{\orgname{Innopolis University}, \orgaddress{\city{Innopolis}, \country{Russia}}}
\affil[2]{\orgname{MB3R Lab}, \orgaddress{\city{Innopolis}, \country{Russia}}}

\abstract{Model-based fault experiments can assess the consequences of component failures before intervening in a running application. Recent work on model discovery and graph simulation acquired operation models from application telemetry and used them for availability prediction. Applying this approach to operational legacy systems raises an acquisition problem: the required traces may be absent or incomplete, while introducing them can require changes across heterogeneous components. These systems continue to serve business operations during extended maintenance and modernization. External observation offers a possible route to model-based fault analysis before a tracing retrofit, but collecting telemetry does not by itself establish that an executable operation model can be obtained. We investigate this route using extended Berkeley Packet Filter (eBPF) observations and the existing graph-based modelling approach. A 108-case study varied application software development kit (SDK) tracing across absent, partial and full conditions in three public applications. Discovered dependencies were combined with declared operation requirements, deployment facts and scenario assumptions; model forecasts preceded independent execution checks. With SDK tracing disabled, the tested OpenTelemetry eBPF Instrumentation path supported all seven operations in two applications across three repetitions. Forecasts agreed exactly in 17 of 21 operation--repetition groups; the maximum state-weighted discrepancy was 4.22 percentage points. Combining observed context links supported nine additional groups under partial tracing. The results establish an acquisition path for virtual fault experiments under the tested observation constraints and identify limits imposed by protocols, evidence association and operation semantics.}

\keywords{model discovery, virtual chaos engineering, runtime models, eBPF, legacy systems, availability modelling}
\maketitle

\section{Introduction}\label{sec:introduction}
A running application may have to remain in service while its operators assess the consequences of component failures. Virtual fault experiments address this need by changing component states in a model and calculating the effects on specified operations. The model provides an analytical setting for examining failure scenarios before deciding which interventions to execute on the application. Its practical use depends on obtaining a model of the deployed system with a meaning that can be checked against application behaviour.

This paper continues a recent research line on model discovery and graph simulation. The initial study recovered service dependencies and replica information from Jaeger observations, then compared graph-based availability forecasts with live fault injection~\cite{krasnovsky2026icse}. A subsequent study examined asynchronous dependencies and immediate operation availability~\cite{krasnovsky2026aina}. The approach represents the dependencies needed by an operation, combines them with replica and execution assumptions, and evaluates the operation under specified failure states. Later work formalized the success event and the conditions for reducing its representation~\cite{krasnovsky2026reduced}; the Procrustes--Bering--Sheaft toolchain implemented acquisition and graph execution~\cite{krasnovsky2026toolchain}. These works provide the modelling basis of the present investigation. Section~\ref{sec:model} explains that basis and the conditional instance used here without requiring prior familiarity with the toolchain.

The demonstrated acquisition paths depended on suitable application traces. That prerequisite restricts their use when a deployed application has no tracing SDK or instrumentation covers only part of an operation. Operational legacy systems are a consequential setting: their components continue to support business functions while compatibility obligations, technological heterogeneity and limited scope for code changes constrain instrumentation~\cite{fritzsch2019,boehm2026,li2022}. Inserting missing trace points and propagating context across service boundaries can become a separate engineering undertaking~\cite{ashok2024,shen2023}. A method that requires the resulting traces leaves model acquisition waiting for that undertaking.

\begin{figure}[t]
\centering
\includegraphics[width=\textwidth]{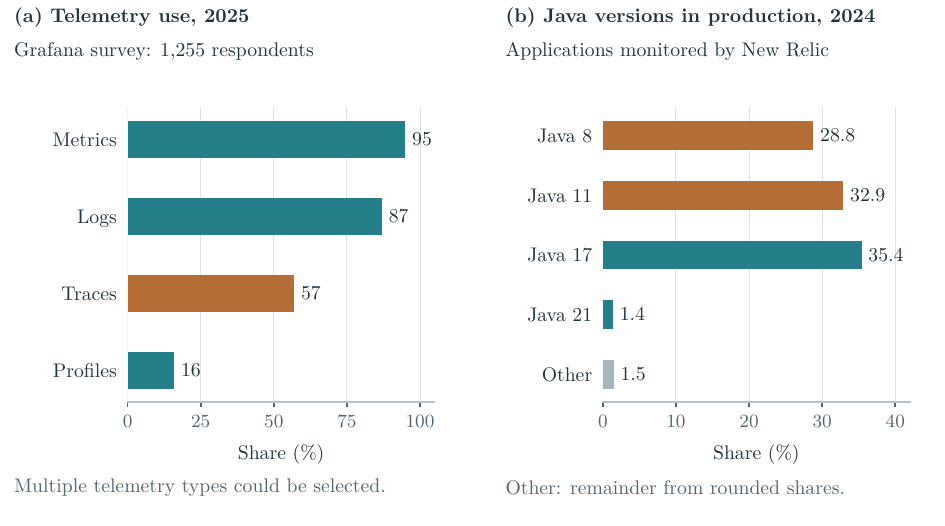}
\caption{Two independent observations motivating the acquisition problem. (a) Types of telemetry used by respondents to Grafana's 2025 survey ($n=1{,}255$; multiple answers permitted)~\cite{grafana2025survey}. The survey records use, not completeness of operation coverage. (b) Java runtime versions in New Relic's 2024 production-application sample~\cite{newrelic2024java}. These applications already reported to an APM service. The 1.5\% ``Other'' category is the remainder from the rounded shares. The cohorts have different selection mechanisms and denominators; their percentages cannot be combined into a prevalence estimate for untraced legacy systems. Runtime version alone does not define legacy status.}
\label{fig:motivation}
\end{figure}

The available industry measurements give two distinct views of this setting (Fig.~\ref{fig:motivation}). Grafana's 2025 survey recorded substantially less use of traces than of metrics or logs among its respondents~\cite{grafana2025survey}. New Relic's 2024 production measurements documented the coexistence of Java runtime generations~\cite{newrelic2024java}. These observations motivate examining the acquisition prerequisite; neither dataset measures the population of untraced legacy applications. Section~\ref{sec:legacy-problem} connects them to empirical studies of continued operation, instrumentation effort and stack compatibility.

External observation provides a technical opportunity to extend model discovery to this setting. DeepFlow, TraceWeaver and subsequent work have investigated obtaining and correlating execution observations without application-level context propagation throughout every service~\cite{shen2023,ashok2024,geng2025deeptrace,hong2026}. Their results establish capabilities for observation and diagnosis. Our analytical objective is to acquire an executable operation model and use it for a virtual fault experiment. The study therefore uses existing collectors and evaluates whether their retained evidence satisfies the requirements of that modelling task.

The research question is whether eBPF observations can extend the existing model-discovery approach to applications with absent or incomplete SDK tracing, and whether models obtained in this way support forecasts consistent with subsequent operation outcomes. We investigate it through 108 experimental cases: three public applications, three SDK conditions, three eBPF pipelines and a no-agent control, with three repetitions. The applications reproduce controlled deficits in tracing; the industrial literature establishes the legacy motivation. The target population is existing applications on a compatible platform whose required observations are unavailable, including legacy and heterogeneous deployments.

To the best of our knowledge, this is the first empirical study of using eBPF observations to extend model discovery for virtual fault experiments to absent or incomplete SDK tracing. The contribution has two parts. First, the comparative acquisition experiment establishes which configured observation paths support the declared operation models as SDK coverage changes. Second, prospective outcome checks and retained source-level evidence examine the resulting models and expose the conditions under which acquisition or prediction fails. With SDK tracing disabled, the tested OpenTelemetry eBPF Instrumentation (OBI) path supported all 21 operation--repetition groups in Petclinic and OpenTelemetry Demo. Observed-context combination additionally supported nine groups under partial tracing. These results extend the demonstrated acquisition conditions of the research line; the model definitions and graph simulator are inherited from the preceding studies.

Section~\ref{sec:legacy-problem} synthesizes the operational problem and technological scope. Section~\ref{sec:model} explains the modelling basis and the studied instance, followed by related work in Section~\ref{sec:related}. Sections~\ref{sec:workflow} and~\ref{sec:design} specify acquisition and experimental design. Section~\ref{sec:results} reports the findings; Sections~\ref{sec:discussion} and~\ref{sec:threats} discuss their implications and limits.

\section{Virtual fault experiments for systems with incomplete tracing}\label{sec:legacy-problem}

The motivating application is a deployed system for which engineers need to assess an operation under component failures while application tracing remains incomplete. Earlier work in this research line acquired dependency graphs from application traces and used graph simulation for such experiments~\cite{krasnovsky2026icse,krasnovsky2026aina,krasnovsky2026toolchain}. Extending that workflow to systems without those traces requires an alternative source of model evidence. Operational legacy systems make this requirement consequential: they retain business responsibilities while inherited technologies and maintenance constraints restrict changes. Projects with limited instrumentation capacity encounter the same acquisition barrier. The literature below establishes why this barrier persists, which costs it entails and which deployed systems an eBPF acquisition path could reach.

\subsection{Operational continuity during extended change}
Reviews of modernization place existing systems at the centre of a continuing engineering process. Hasan et al.\ synthesized 47 studies of legacy-to-cloud migration, covering activities before, during and after migration and the technical and organizational constraints of systems supporting core business processes~\cite{hasan2023}. Wolfart et al.'s mapping of 62 studies identified understanding the existing system and integrating old and new components as parts of microservice modernization~\cite{wolfart2021}. Their roadmap included execution traces, call graphs and usage scenarios among the representations needed to understand a system before transformation.

Industrial cases document the continued production responsibilities behind that need. Fritzsch et al.\ interviewed 16 practitioners about 14 systems in ten companies~\cite{fritzsch2019}. Five cases ran old and new systems in parallel. One retained an SAP Hybris shop in production while its replacement was developed; another concerned a public-transport management system that had evolved for almost 30 years. Reported migration periods ranged from 1.5 to more than three years, including both completed durations and estimates for ongoing projects. A 2026 case studied a European port operator's terminal system, developed since 2006 and still used for gate processing, cargo tracking and billing~\cite{boehm2026}. Its older .NET Framework, ASP.NET MVC and Web Forms components constrained incremental replacement.

Michael Ayas et al.\ examined this coexistence through interviews with 19 practitioners from 16 companies and 215 StackOverflow posts comprising 76 selected questions and their answers~\cite{ayas2023}. Teams had to maintain the old system while extracting services; an API layer around it introduced additional development and maintenance. These studies establish a practical interval during which fault-related decisions concern the architecture currently in service. Model acquisition deferred until modernization is complete would omit that interval. They document operational persistence in studied organizations without estimating a universal share of enterprise systems classified as legacy.

\subsection{Tracing as an engineering prerequisite for model acquisition}
Observation coverage can remain uneven across a maintained system. Usman et al.'s survey described the integration of heterogeneous infrastructure and applications originally designed without observability requirements~\cite{usman2022}. Giamattei et al.'s systematic grey-literature review of 71 monitoring tools treated probes and tracing code as artifacts requiring development and maintenance~\cite{giamattei2024}. A trace-dependent model consumer inherits these requirements: the relevant components must expose compatible observations, and interactions must remain associated across service boundaries.

Li et al.'s industrial study examined ten companies, eight of whose studied systems used distributed tracing~\cite{li2022}. The three largest systems supplemented framework or dynamic-binary instrumentation with manual code, partly because available mechanisms did not cover legacy services. Difficult-to-instrument services also contributed to incomplete tracing. TraceWeaver identified cross-team changes and legacy or proprietary components as obstacles to propagating trace context~\cite{ashok2024}. Such gaps matter directly to an operation model: a missing association can prevent the representation of a required call even though the application executes it successfully.

DeepFlow provides a concrete effort observation. Six of its ten surveyed commercial customers reported needing hours or days to instrument a single component~\cite{shen2023}. Its production evaluation also described external observation of a running service with incomplete tracing, without application-code changes. The reported effort concerns selected customers and component instrumentation. It supports the existence of a material integration burden; it does not quantify the cost of retrofitting an entire enterprise application or modernizing its architecture.

Engineering effort and runtime overhead answer different questions. The former includes selecting compatible instrumentation, adding missing propagation, validating observations and maintaining them through changes. Telemetry processing and storage impose separate operational costs~\cite{giamattei2024}. Automatic instrumentation already reduces application changes for supported libraries and runtimes. The acquisition problem therefore depends on the deployed stack and required evidence: full migration or refactoring is not a universal prerequisite for tracing, while completing the necessary coverage can still require substantial work. An external acquisition path could bring model-based fault experiments forward in that process.

\subsection{Runtime diversity and the boundary of external acquisition}
Figure~\ref{fig:motivation} presents two complementary observations. Grafana's 2025 survey collected 1,255 responses through its community and industry events; 95\% reported metrics, 87\% logs and 57\% traces~\cite{grafana2025survey}. New Relic's 2024 report described runtime versions in production applications already monitored by its agents~\cite{newrelic2024java}. The survey records reported signal use, while the production cohort records Java versions. Neither measures complete tracing coverage, and their samples cannot be combined to estimate how many legacy applications lack tracing. Together they motivate examining both observation availability and runtime diversity, with their populations kept separate.

Research on software stacks explains why compatibility deserves explicit treatment. Petrulio et al.\ analysed 50 Java APIs, interviewed 23 API producers and surveyed 110 developers~\cite{petrulio2021}. Backward compatibility and enterprise clients remaining on Java 7 constrained adoption of newer language features in that historical sample. Sul\'ir et al.\ subsequently attempted to build 2,500 GitHub projects with JDK versions 6--23~\cite{sulir2026}. From Java 8 onward, success rates ranged from 30 to 44\%; downgrading the latest JDK repaired 52\% of its failed builds. These results concern local buildability, which identifies a concrete obstacle to changing or rebuilding existing code. They provide no production adoption estimate. Lubas et al.'s study of 553 microservice applications found 24 implementation languages~\cite{lubas2025}. Its repository population documents diversity and benchmark selection constraints rather than the composition of enterprise deployments.

Application age, runtime compatibility and observation support consequently require separate checks. OBI targets Linux and imposes kernel, privilege and protocol requirements~\cite{obiDocumentation}. Microsoft's .NET Framework runs on Windows, whereas modern .NET has a cross-platform implementation~\cite{microsoftNetFramework}. The port-system case therefore substantiates legacy production use without establishing applicability of the tested Linux collection path to that system. More generally, old Java or .NET code cannot be classified as observable from its age or language alone. The deployed operating system, probe availability, communication protocol and collector implementation determine which evidence can be acquired. The experiment examines Linux deployments with explicitly assessed protocol support. Its results retain those platform and collector constraints.

\subsection{The modelling problem and its empirical scope}
These findings identify an acquisition problem within the preceding virtual-chaos-engineering workflow. Existing systems may require fault analysis during prolonged maintenance or modernization, while their application traces omit dependencies required by the model. External observation has already addressed parts of this telemetry deficit~\cite{shen2023,ashok2024}. The research question here concerns extending the model-based workflow to that setting: can eBPF observations supply sufficient structural evidence to construct the declared operation model, and do its frozen forecasts agree with subsequent operation outcomes?

The required evidence exceeds a list of communicating addresses. The model needs logical identities and observed dependencies that can be matched to operation requirements. Operation success, mandatory completion, replica placement and the failure-scenario law remain explicit inputs. In particular, a response acknowledgement and a durable write define different success events. Collecting a service map does not settle those semantics. Section~\ref{sec:model} defines the inherited model and the conditional instance evaluated here.

Working with the deployed application consequently means acquiring model evidence before completing application SDK trace recording and export. It still requires deploying and configuring observation components. The tested OBI path used active context propagation; the absent-SDK condition disabled recording and export without removing every SDK library from application binaries. These conditions delimit the demonstrated preparation burden.

The three public applications supply controlled contrasts in tracing coverage and protocols. They test the acquisition premise relevant to legacy and underinstrumented deployments, while production persistence and integration costs come from the literature above. The study measures model support and later forecast agreement under common supplied contracts and fault assumptions. Successful acquisition demonstrates that the tested operation model could be constructed before complete SDK tracing. Unsupported cases and forecast discrepancies identify limits of that extension. Developer-time savings and deployment costs were outside the measured outcomes.

\section{Models for virtual fault experiments}
\label{sec:model}

\subsection{Purpose and representation}
We use \emph{virtual chaos engineering} to denote experiments with declared fault scenarios on an executable model of a running application. A model experiment changes the represented state of components or replicas and calculates the consequences for a specified operation. The earlier model-discovery and graph-simulation studies developed this use of application observations~\cite{krasnovsky2026icse,krasnovsky2026aina}; the present study examines its acquisition premise when the application supplies insufficient SDK tracing. Physical interventions in the later validation runs check the model's forecasts.

The representation connects three kinds of information. Observations establish service identities and interaction evidence. An operation contract specifies the entry and the calls required for the chosen meaning of success. Deployment and scenario declarations describe replicas and the fault states to examine. The resulting model retains the observed dependencies to which those declarations apply. Graph simulation evaluates whether the represented operation requirements remain satisfied in each declared state. Scenario weights then aggregate these state-specific answers into a conditional success probability.

\begin{figure}[t]
\centering
\includegraphics[width=\linewidth]{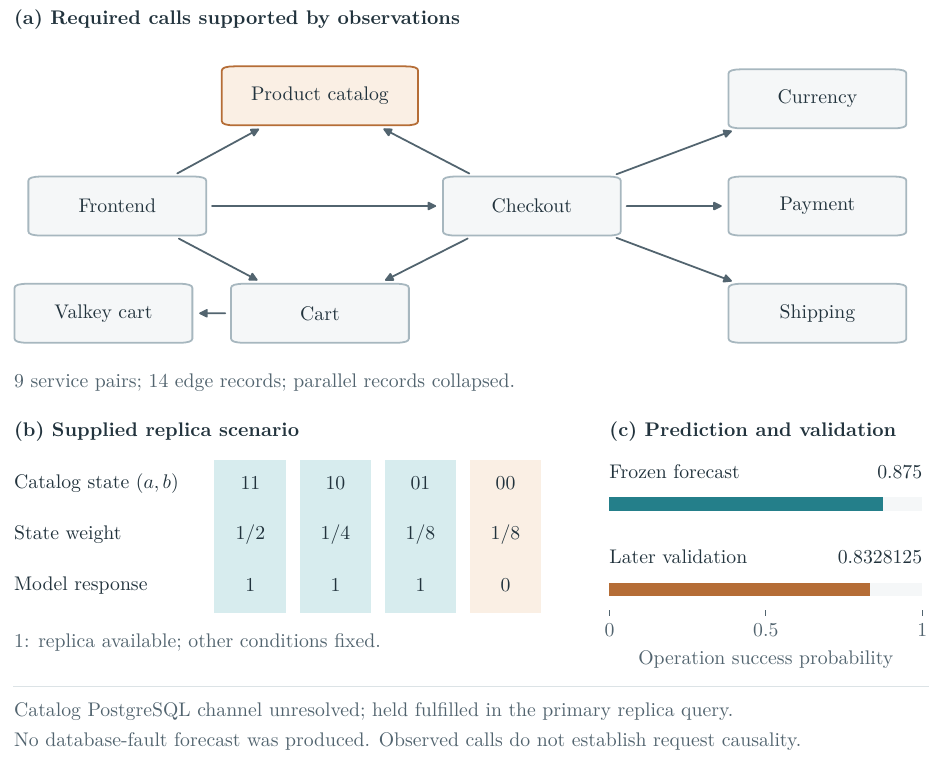}
\caption{A model for a virtual fault experiment: the saved OpenTelemetry Demo checkout case with SDK tracing absent, OBI, repetition~9. The required observed subgraph contains nine service pairs and fourteen edge records; parallel records are collapsed. Replica facts, fulfilled conditions and scenario weights are supplied declarations. The model forecast precedes the independent application outcomes. The Product Catalog PostgreSQL channel remained unresolved: the primary query conditions on its fulfilment, and the database-fault forecast is unavailable. The graph omits unrelated source-global dependencies and does not represent a reconstructed per-request checkout trace.}
\label{fig:worked-model}
\end{figure}

Figure~\ref{fig:worked-model} illustrates this representation with the saved checkout example from OpenTelemetry Demo. Its admitted dependency structure contains nine service pairs represented by fourteen required edge records. The intervention controls the two Product Catalog replicas. Under the supplied conditions, the model predicts success whenever at least one of them remains live. The four state responses, combined with the declared weights, give a forecast of $0.875$. Subsequent application outcomes test this conditional abstraction; Section~\ref{sec:results} reports their agreement and discrepancies. The acquired graph and the supplied operation semantics are therefore both necessary to interpret the forecast.

\subsection{Operation semantics and evidence}
Availability concerns readiness for correct service, whereas reliability concerns continuity of correct service~\cite{avizienis2004}. The quantity studied here is the probability of a specified operation-success event under a declared execution-state law. An operation model therefore requires both a success contract and evidence that supports its representation. Krasnovsky and Maslovskaya~\cite{krasnovsky2026reduced} formalize an operation-specific stochastic model
\begin{equation}
  \mathcal M_u=\langle G_u,\mathcal R,P,\Phi_u\rangle,
\end{equation}
where $G_u$ is a typed service-dependency graph, $\mathcal R$ describes replicas, $\Phi_u$ specifies operation requirements and observation bindings, and $P$ is a joint execution-state law. Their execution event combines structural reachability, delivery to the selected replicas, completion of mandatory logical call groups, and deadline satisfaction:
\begin{equation}
 E_u(W)=B_u(X)\land\bigwedge_s J_s(X,D)
         \land H_G\land\bigwedge_j C_j\land T,
 \qquad A_u^{\mathcal M}=\mathbb E_P[E_u(W)].
 \label{eq:prior-execution}
\end{equation}
Here $X$ contains eligibility coordinates, $D$ records replica selections, $C_j$ denotes completion of a mandatory logical group, and $T$ denotes deadline satisfaction. $B_u$ and $H_G$ express reachability and required structural links. Correspondence with externally checked application success depends on the operation contract and the evidence bindings.

That framework also separates unsupported structure from incomplete observations of a supported representation. A masked observation $o$ admits compatible states $\mathcal F(o)$; joint laws consistent with those states yield attainable probability bounds. Its reduction results identify assumptions under which completion-and-deadline information suffices for a smaller target representation~\cite{krasnovsky2026reduced}. The present study uses a fixed conditional structural instance of that framework. It tests acquisition and forecast adequacy for this instance; the general masked-state identification procedure remains outside the implementation evaluated here.

\subsection{Discovered structure and supplied declarations}
Let $\widehat G=(\widehat V,\widehat E,\tau)$ be the graph recovered by Bering from a frozen observation view. Let $\Gamma_u$ contain the operation contract and its graph-admission requirements, and let $\Theta$ contain supplied service identities and replica-placement facts. Binding compiles those declarations against observed structure:
\begin{equation}
 \operatorname{Bind}(\widehat G,\Gamma_u,\Theta)
   \in \{(v_u,V_u^{\rm req},E_u^{\rm req})\}\cup\{\bot\}.
 \label{eq:binding}
\end{equation}
A supported result identifies an observed entry $v_u$, required targets $V_u^{\rm req}$, and synchronous edges $E_u^{\rm req}$ under the chosen contract. The symbol $\bot$ denotes unavailable support. Supplied aliases normalize observed identities; operation metadata attaches predicates and replica facts to existing graph entities. Neither operation declarations nor aliases insert missing dependency edges or establish execution ancestry.

Here $\operatorname{Bind}$ denotes the existing contract-compilation step of the toolchain~\cite{krasnovsky2026toolchain}. Section~\ref{sec:workflow} specifies its admission rules for the source-global graph; the present study evaluates the evidence supplied to those rules.

The admission requirements are specific to the chosen abstraction. In particular, the corrected Social Network acknowledgement model retains a conservative normal-path requirement when admitting a graph for the replica-intervention analysis. That list is not an unconditional conjunction defining HTTP acknowledgement in every failure mode. A different acknowledgement model could admit different observations; an unavailable binding here establishes failure of the tested acquisition-and-binding path under its declared requirements.

The primary analysis conditions on fulfilment of nonintervened requirements. A database channel absent from the observed graph can therefore remain an explicitly supplied fulfilled condition when the required application path is supported. Such a condition is not counted as a recovered dependency. A database-down query changes that scope and requires the applicable conditional contract, evidence for the selected channel, and its fault-group attribution. Primary and database-fault support are evaluated separately.

\subsection{Conditional structural availability}
The primary intervention controls two replicas of one service. Let $x=(x_a,x_b)\in\{0,1\}^2$ denote their live states. A controlled service is structurally live when at least one replica is live. Nonintervened service and edge guards remain true within the declared scope. Let $\widehat E_x$ contain synchronous edges with live endpoints, and let $\operatorname{Reach}_x(v_u)$ contain live services reachable from the entry. The implemented structural event is
\begin{align}
 \psi_u(x)={}&
 \mathbf 1\{v_u\in\operatorname{Reach}_x(v_u)\}
 \prod_{v\in V_u^{\rm req}}\mathbf 1\{v\in\operatorname{Reach}_x(v_u)\}
 \nonumber\\[-2pt]
 &\cdot\prod_{e\in E_u^{\rm req}}
 \mathbf 1\{e\in\widehat E_x,
              \operatorname{src}(e)\in\operatorname{Reach}_x(v_u)\}.
 \label{eq:structural-event}
\end{align}
The supplied joint intervention law is
\begin{equation}
 (\pi_{11},\pi_{10},\pi_{01},\pi_{00})
   =\left(\tfrac12,\tfrac14,\tfrac18,\tfrac18\right).
 \label{eq:intervention-law}
\end{equation}
These weights specify a scenario mixture. They are not estimated from target outcomes and do not assume independent replica states. For a supported binding, the forecast is
\begin{equation}
 a_u(\widehat G,\Gamma_u,\Theta)
   =\sum_{x\in\{0,1\}^2}\pi_x\psi_u(x).
 \label{eq:conditional-availability}
\end{equation}
Sheaft evaluates each deterministic profile on the saved model. The workflow checks its result against an exact graph-level evaluation of Eq.~\eqref{eq:structural-event}, then applies the weights. The configured trial count is 100,000 per profile; fixed guards give identical structural outcomes on repeated trials, so these forecasts have no Monte Carlo variation. State vectors $(1,1,1,0)$ and $(1,1,1,1)$ produce $7/8$ and one, respectively. The structural rule abstracts replica selection, timing, pool behaviour and additional runtime failures into fulfilled conditions. External operation outcomes test whether that abstraction was adequate in the executed scenarios.

Let $Z$ collect these fulfilled conditions. The experiment fixes some components of $Z$ by configuration and checks selected postconditions independently; it does not verify every component for every target attempt (Table~\ref{tab:model-conditions}). Comparing the conditional forecast with all observed target outcomes therefore tests the joint adequacy of the acquired structure and supplied execution assumptions. A discrepancy alone does not identify which component failed.

\subsection{Support, baselines and validation}
An unsupported binding produces no point forecast. Its recorded range $[0,1]$ denotes unavailable information, rather than an application of the preceding paper's sharp-bound algorithm. A supported forecast has a singleton range. Neither type of range is a statistical confidence interval.

The direct conditional reference receives the same operation semantics, controlled two-replica inventory and scenario law. It evaluates the dependence declared by the contract without recovering a graph or fitting validation outcomes. Equality with this reference checks the numerical consequence of the model's supplied conditions. The acquisition experiment asks whether the available observations support an explicit dependency representation that can be bound to those operation requirements and executed for the declared fault scenarios. Forecast agreement with later application outcomes is a separate test of that representation and its execution assumptions. The observed-context union, defined in Section~\ref{sec:workflow}, additionally tests whether straightforward combination of the available observations resolves acquisition failures.

\section{Related work and research positioning}\label{sec:related}

The acquisition premise of the preceding operation-model workflow links this study to quantitative model extraction and to external observation. The comparison concerns the target model, required observations and subsequent fault experiments. Janes et al.'s multivocal review of 30 tracing tools documented variation in supported technologies and capabilities~\cite{janes2023}; model acquisition additionally depends on which of those capabilities supplies evidence for the chosen analysis.

\subsection{Model acquisition, Palladio and virtual fault experiments}
Models@run.time relate an executing system to abstractions used for reasoning~\cite{bencomo2019}. Architectural reconstruction established interpretation rules between implementation evidence and model elements: DiscoTect used event mappings expressed with Colored Petri Nets~\cite{schmerl2006}; work on existing ROS systems combined static extraction and monitoring~\cite{garcia2021}. MicroART, Mayer and Weinreich's approach, Retriever and MiSAR recovered architectural views from combinations of development, deployment and runtime evidence~\cite{granchelli2017,mayer2018,kirschner2025,alshuqayran2025}. These recovered views supply structure; quantitative fault analysis requires additional semantics and assumptions.

Palladio provides a substantive predecessor for observation-based quantitative modelling. Architecture-level performance extraction~\cite{brosig2011} and the Performance Model Extractor (PMX)~\cite{walter2017} recovered models describing components, calls, deployment and resource-related behaviour. The published PMX extraction service required Kieker traces. For accurate resource-demand estimation, the traces had to come from a low-load run or be supplemented with resource-utilization measurements~\cite{walter2017meaas}. Its target included resource and workload information needed for performance predictions. WESSBAS recovered workload models from session logs~\cite{vogele2018}; online learning~\cite{spinner2019}, iObserve~\cite{heinrich2020} and CIPM~\cite{mazkatli2025} addressed adaptation or calibration as software and workloads changed. These approaches expose the dependence of model inputs on the intended analysis.

Palladio's reliability extension represented control flow and component failures~\cite{brosch2012}. More recently, Weber et al.\ extended PMX to construct Palladio performability models from OpenTelemetry spans and errors~\cite{weber2025}. Their component mapping was tailored to Spring Boot auto-instrumentation. The Petclinic demonstration used an OpenTelemetry agent and Jaeger; fault injection checked pipeline functionality and error representation, while prediction-accuracy assessment remained subsequent work. eBPF-derived records could contribute to such extraction if they supplied the required fields and mappings. The published pipeline therefore establishes an acquisition dependency in a particular implementation, without imposing a fundamental incompatibility between Palladio and eBPF.

Other dependability methods clarify the distinction between recovered structure and quantitative assumptions. Architectural reliability surveys and simulation~\cite{immonen2008,gokhale2005}, Bayesian runtime updating~\cite{epifani2009} and probabilistic quality-of-service management~\cite{calinescu2011} used different model coordinates. Hora combined architectural dependencies, per-request traces and resource measurements with component predictors in a Bayesian failure-propagation model~\cite{pitakrat2018}. MiSim simulated transient resilience from supplied architecture, operation demands, dependency probabilities, mechanisms, workload and faultload~\cite{frank2022misim}. Studies of Retry and Circuit Breaker patterns further demonstrated the effect of execution mechanisms on outcomes~\cite{mendonca2020,aderaldo2025}.

The present analytical target is the operation model developed in the preceding studies and defined in Section~\ref{sec:model}. Its studied instance combines acquired dependencies with declared mandatory calls, replica placement and a supplied state law. Virtual fault experiments evaluate the specified operation-success event under that law. This scope explains which evidence is needed from eBPF and which performance-model parameters are outside the experiment. The comparison concerns extension of that particular workflow under missing traces; it does not establish a general ranking of modelling formalisms.

\subsection{External observation and eBPF-derived models}
External inference preceded current eBPF tooling. X-Trace and Dapper associated activities through propagated metadata~\cite{fonseca2007,sigelman2010}. Message-based causal inference~\cite{aguilera2003}, Magpie's event schemas~\cite{barham2004}, WAP5's probabilistic paths~\cite{reynolds2006} and vPath's thread/network correlations~\cite{tak2009} supplied other association mechanisms. Sherlock and Orion examined dependencies for diagnosis and the limits of network evidence~\cite{bahl2007,chen2008}. This literature separates communication evidence, inferred causality and application semantics.

\paragraph{Measurements and dependencies.}
Cassagnes et al.\ profiled containerized applications without requiring their own performance metrics~\cite{cassagnes2020}; ViperProbe used eBPF for adaptive metric collection~\cite{levin2020}. Socket tracing recovered service dependencies from TCP/UDP activity~\cite{tsubouchi2022}, while cross-domain observation and netobserv addressed connection measurements and network flows~\cite{ranjitha2022,kannan2024}. Request-level approaches progressed from syscall proxies for throughput and saturation~\cite{rezvani2024} to request-boundary reconstruction for latency and throughput~\cite{ibnath2026}. These outputs expose different aspects of execution. A communication edge alone does not state whether an operation requires that call to complete.

\paragraph{Context and causal association.}
DeepFlow introduced a network-centric tracing plane with implicit context propagation~\cite{shen2023}. ZeroTracer generated traces in the kernel with in-band HTTP propagation~\cite{yang2025}; ChainScope extended kernel-level propagation and sampling to more complex concurrency and remote-call patterns~\cite{hong2026}. Their results establish prior achievements in obtaining traces without application SDK code, including production experience in DeepFlow.

TraceWeaver reconstructed paths using timing and call-graph information from test environments~\cite{ashok2024}. DeepTrace associated eBPF-derived spans through transaction content, endpoint correlations and timing~\cite{geng2025deeptrace}. FlowScope estimated method-pair delays and selected associations through constrained inference~\cite{geng2026flowscope}. XPLOG combined application log writes and syscall events into causally ordered cross-host streams~\cite{satapathy2026}. Such methods can recover associations that are absent from explicitly propagated context. Our evaluated codec retains observed identifiers; statistical parentage inference was outside its implementation. Accordingly, an unsupported model in the tested export path does not establish a limitation of every reconstruction method reported for that tool or protocol.

\paragraph{Target-specific model evidence.}
LiME extracted Linux real-time task models from eBPF observations for schedulability analysis~\cite{brandenburg2025}. Liegsalz and Brunnert's eBPF jAgent combined eBPF and HotSpot USDT probes to obtain transaction-level resource demands from Java applications and export them through OpenTelemetry~\cite{liegsalz2026jagent}. The authors explicitly connected those measurements with performance-model parameterization. eInfer reconstructed execution graphs for distributed language-model inference using kernel, userspace and accelerator-facing events~\cite{chu2025}. An additional preprint used dependency discovery to guide microservice migration~\cite{trivedi2026}. These studies establish that eBPF-derived observations already served analytical targets. The present target is an operation model for virtual fault experiments when its earlier application-trace inputs are unavailable or incomplete.

Protocol parsing and implementation fidelity delimit each acquisition path~\cite{kumar2024,machado2025}. Published overhead evaluations also use different probes, workloads and collection objectives: DeepTrace compared tracing configurations~\cite{geng2025deeptrace}, whereas jAgent measured transaction-level resource tracing~\cite{liegsalz2026jagent}. Those measurements provide context for observation costs without establishing an overhead advantage for the model-acquisition paths evaluated here.

\begin{table}[tbp]
\caption{Representative analytical targets and acquisition requirements. Requirements describe the cited implementations and the studied model instance.}
\label{tab:related-approaches}
\centering\small
\begin{tabularx}{\linewidth}{@{}>{\raggedright\arraybackslash}p{31mm}>{\raggedright\arraybackslash}p{43mm}>{\raggedright\arraybackslash}X@{}}
\toprule
Approach & Analytical object & Evidence and supplied information\\
\midrule
PMX/Palladio~\cite{walter2017meaas,weber2025} & Architectural performance or performability model & Execution traces with component mappings; resource or error observations according to the target analysis\\
\addlinespace
DeepFlow; ZeroTracer; ChainScope~\cite{shen2023,yang2025,hong2026} & Correlated request traces & Supported protocols and execution-context propagation mechanisms\\
\addlinespace
TraceWeaver; DeepTrace; FlowScope~\cite{ashok2024,geng2025deeptrace,geng2026flowscope} & Reconstructed request paths & Timing, content or prior call-graph evidence and approach-specific inference rules\\
\addlinespace
LiME; eBPF jAgent~\cite{brandenburg2025,liegsalz2026jagent} & Real-time task models; transaction resource demands & Task events or Java runtime probes selected for the analytical target\\
\addlinespace
Present study & Operation model for virtual fault experiments & Retained dependencies plus declared operation requirements, identities, deployment and state law; later outcomes validate frozen forecasts\\
\bottomrule
\end{tabularx}
\end{table}

The comparison in Table~\ref{tab:related-approaches} identifies the modelling question left open by successful collection: whether the retained evidence instantiates the intended operation model. The experiment tests this under absent, partial and full SDK tracing, with shared operation contracts and fault assumptions. Native SDK evidence, the union of observed context links and direct conditional calculation distinguish acquisition support from prediction arithmetic. The measured outcomes are supported construction and subsequent forecast agreement.

\subsection{Fault-based testing and validation}
Gremlin translated supplied graphs and outage recipes into interventions on running applications~\cite{heorhiadi2016}. Filibuster expanded functional tests into service-failure scenarios using instrumentation and execution feedback~\cite{meiklejohn2021}. Lineage-driven fault injection selected failure hypotheses from successful-execution provenance~\cite{alvaro2015}; its Netflix adaptation accumulated alternative call graphs through trace-guided tests~\cite{alvaro2016}. These approaches exercised application faults directly. Virtual experiments in our preceding workflow execute on the model, followed here by separate application interventions to validate the frozen forecasts.

EMART combined monitored usage and demand outcomes with adaptive testing for operational reliability estimation~\cite{pietrantuono2020}. Our later outcomes serve only as validation evidence: they neither fit the supplied state law nor revise the frozen graphs. This design follows Sargent's separation of conceptual-model validity, implementation verification and operational validation~\cite{sargent2013}. Forecast agreement remains evidence for the tested abstraction and scenarios, with its structural-identification limits stated separately.

\subsection{Relationship to the preceding research line}
The first study acquired a connectivity model from Jaeger observations of DeathStarBench Social Network and compared graph simulation with live fault injection~\cite{krasnovsky2026icse}. A second examined asynchronous dependencies in OpenTelemetry Demo~\cite{krasnovsky2026aina}. Procrustes--Bering--Sheaft already compiled declared mandatory calls into predicates over discovered edges~\cite{krasnovsky2026toolchain}. The subsequent stochastic formulation separated structure, operation semantics and state laws and treated identification under incomplete evidence~\cite{krasnovsky2026reduced}. These contributions supply the model and implementation used here.

The new evidence tests extension of that workflow beyond complete application tracing through a blocked 108-case study of three applications, three SDK conditions, four runtime treatments and three repetitions. The corrected analysis retains 72 Petclinic/OTel validation records and uses fresh outcomes for 36 Social Network configurations with corrected contracts and the original frozen graphs. Earlier Social Network outcomes document the semantic correction rather than additional repetitions. The experiment contributes acquisition and validation evidence for the declared conditional model; quantitative results from preceding campaigns remain separate.

\section{From observations to an executable model}\label{sec:workflow}

The acquisition workflow uses existing observation and modelling tools to construct the representation defined in Section~\ref{sec:model}. Its outputs record the admitted dependencies, the operation requirements attached to them and the executable fault scenarios. Closing the observation input and freezing each forecast before the validation deployment separates model acquisition from the subsequent check of predicted behaviour.

\subsection{Runtime treatments and observation sources}
Each experimental case fixes an application, SDK-coverage mode, repetition and runtime treatment. The treatments are no study eBPF agent, OBI, Coroot and DeepFlow. Each application deployment spans two Linux guests with separate kernels on one GitHub Actions worker. Collectors run in separate treatments. The no-agent treatment supplies a tracing-disabled control with absent SDK tracing and clean native comparisons with partial or full SDK tracing.

Within each case, four methods consume native SDK input, eBPF input, the combined input, or an observed-context union. These are analyses of the same case, not four additional deployments. In an agent treatment, the native method denotes its SDK projection and can therefore differ from the clean no-agent control. Empty source inputs remain explicit. OBI uses configured active context propagation; the evaluated treatment does not represent passive kernel observation alone.

Collector-specific codecs convert retained exports into Bering inputs while preserving source identities, provenance locators and observed trace/span/parent identifiers. They do not create distributed parentage from timestamps, process identifiers or shared addresses. Supplied guest-scoped identities and deployment labels associate observed services with declared roles. Replica inventory and placement are supplied facts; assigning two replica slots does not demonstrate discovery of physical replica identity.

\subsection{Operation entry and structural evidence}
The workload driver supplies attempt boundaries and an opaque request identifier. Entry matching uses an actually observed header, method/path, server role and verified service identity. The same policy applies to native and eBPF observations. It can admit an entry without an exporter-generated span identifier; genuine observed identifiers remain necessary for descendant traversal. A unique observed service/method/canonical-route match can bind an endpoint when request-linked proof is unavailable. These proofs do not filter the discovery input.

Operation-family fitting uses frozen method, route, step and source signatures with observed call-set descriptors. It excludes workload operation labels and external success outcomes. Optional or repeated descendants do not automatically split a known entry family; conflicting, ambiguous and unseen structures retain uncertainty. Family agreement is descriptive and does not establish application success or complete execution semantics.

Bering consumes the complete frozen source prefix, including preparation, verification and background/control observations. Binding matches supplied operation obligations to admitted paths in this source-global graph. A recovered channel is evidence of an observed interaction; it does not establish that every target request traversed the channel. Missing or ambiguous required application paths yield unavailable forecasts. A missing database observation can remain a declared fulfilled primary condition, while a database-down forecast additionally requires channel and fault attribution.

\paragraph{Implemented binding procedure.}
The inputs are the recovered graph and evidence annotations, the declared operation steps and mandatory calls, admitted identity/proxy declarations, and the controlled-service and database-fault mappings. The implementation applies the following rules without using validation outcomes:
\begin{enumerate}
\item Reject an empty frozen input; otherwise resolve declared service names against observed identities and retain identity collisions as uncertainty. Where an independently supplied identity group admits several entry representations, retain the candidates and select in declared order only after their supported deterministic state vectors and known database-fault values agree.
\item For each mandatory application call, resolve its caller and callee and require the caller to be reachable from the entry. Match any supplied call hint by exact RPC tokens or HTTP method and canonical route. Prefer an observed direct synchronous channel; otherwise require one admissible service path through declared transparent proxies. Retain every matching typed edge on the selected hops.
\item Resolve database endpoints from their recorded database evidence and supplied whole-endpoint fault group. A missing or ambiguous channel from a reachable caller may remain an explicit fulfilled primary condition. A forecast for the selected database fault additionally requires its observed channel and fault-group attribution.
\item Bind each declared entry step using its recorded endpoint proof or a unique observed service/method/canonical-route match. A missing or ambiguous endpoint remains an unresolved obligation.
\item Preserve the resolved edge identifiers, targets, endpoint proofs, assumptions and refusal reasons. Unresolved admission requirements leave the primary forecast unavailable; controlled-service evidence is required only when that service occurs in the declared obligations. A supported binding supplies the existing structural predicate and declared fault profiles to Sheaft.
\end{enumerate}
This rule admits an otherwise qualifying dependency observed elsewhere in the frozen source prefix. Support therefore establishes compatibility with the supplied operation obligations at source-global scope; it does not establish that all required edges co-occurred in one operation execution. The study did not estimate edge precision or recall against an independent operation-specific graph reference, or compare this support with a prefix excluding preparation and background traffic.

\subsection{Union of observed context edges}
A child observation can refer to a parent exported by another source. Source namespaces preserve provenance but can leave that cross-source relation unavailable to a single projection. The context baseline tests a specified union of observed evidence under the same contracts and semantic information as the other methods.

The method retains the canonical combined graph and all existing edge records and endpoints. A second pass of the same pinned Bering consumes the same unmodified mixed OTLP bodies through the ordinary OTLP endpoint. The union adds only absent SERVER-child edges satisfying Bering's unambiguous parent/child rule, with both endpoint services already present in the canonical graph. It adds no services, endpoints, CLIENT-fallback edges or inferred physical identities. Conflicts remain explicit uncertainty.

Both original snapshots and the derivation identify this result as a derived graph. Its operation binding and Sheaft calculation use the same implementation and scenario law as the other methods. Reuse of an unchanged union refers to the original forecast, including its unavailable and database results; it does not add a model execution or independent observation. Additional discovery and simulation work remains in the cost inventory. The baseline makes no claim of algorithmic novelty or reconstructed per-request causality.

\subsection{Freezing, execution and provenance}
After capture, the workflow closes and hashes the inputs, then stops the application, collector and guests before offline model construction. Supported operations retain a discovered graph, operation binding, predicate, fault definitions, simulator command and report. All forecasts for the four runtime treatments of a logical block are frozen and published before its validation targets start. Validation uses fresh deployments from the same sealed blueprint.

Each forecast records support, source, operation and graph view. Both the unprojected raw view and the declared network-scope view remain in the analytical records. Missing support retains its reason and no point prediction. Input hashes and producing source/run/job/attempt identities distinguish original capture, model processing and later target evaluation. A technical worker failure can trigger processing of the same closed input; forecast accuracy is not an input-selection criterion.

The compact publication artifact contains source code, analytical results for the complete case selection, provenance and costs, and one selected real capture/model example~\cite{ebpfartifact2026}. It supports reaggregation of the saved results and new executions from the code. Its inclusion of the selected example is not a claim that all historical raw inputs are present or that packaging reran model discovery. Individual latency values and the complete historical classification assignments are outside the compact analytical data.

\section{Experimental design}\label{sec:design}

\subsection{Research questions and experimental matrix}
RQ1 asks whether eBPF observations without application SDK tracing support executable operation models and forecasts checked against subsequent application outcomes. RQ2 asks what eBPF contributes under partial tracing, including the observed-context union. RQ3 asks which protocol, evidence and success-semantics requirements delimit the supported models.

The V6 study contains 108 cases in 27 logical blocks. Each block fixes one of three applications, an SDK mode (absent, partial or full), and a repetition identifier (9, 10 or 11); it contains no-agent, OBI, Coroot and DeepFlow runtime treatments. Four observation methods and two graph views share each case's target outcomes. They are not independent experimental repetitions. Supplementary Material (Online Resource 1) provides the case selection, application and tool revisions, contracts and retained provenance.

Petclinic represents a Java HTTP application backed by MySQL. OpenTelemetry Demo provides heterogeneous HTTP/gRPC paths and asynchronous checkout persistence. DeathStarBench Social Network uses framed binary Thrift calls~\cite{gan2019}. The controlled services are Visits, Product Catalog and User Timeline, respectively, with one controlled replica on each guest. All ten declared operations remain in the analysis (Table~\ref{tab:operations}). A workload operation can contain multiple HTTP steps, so operation attempts and individual HTTP requests are distinct units.

\begin{table}[t]
\caption{Operations, independently checked primary outcomes, and direct conditional forecasts under Eq.~\eqref{eq:intervention-law}. An available graph forecast evaluates the same declared replica-intervention law.}
\label{tab:operations}
\centering\small
\begin{tabularx}{\linewidth}{@{}>{\raggedright\arraybackslash}p{25mm}>{\raggedright\arraybackslash}p{34mm}>{\raggedright\arraybackslash}Xr@{}}
\toprule
Application & Operation & Primary success check & Direct forecast\\
\midrule
Petclinic & List owners & HTTP 200 and canonical payload matching SQL-derived fixture & 1\\
& List vets & HTTP 200 and canonical payload matching SQL-derived fixture & 1\\
& Owner details with visits & Complete owner/pet/visit payload; incomplete fallback rejected & .875\\
& Create visit & HTTP 201, matching marker and exact SQL-row verification & .875\\
\addlinespace
OTel Demo & Browse product & HTTP 200 and matching product identity & .875\\
& Add to cart & Product/cart steps succeed; selected item at quantity one & .875\\
& Checkout & Required steps succeed; nonempty order identifier and matching item & .875\\
\addlinespace
Social Network & Compose post & HTTP success and expected acknowledgement marker & .875\\
& Read home timeline & HTTP success and stable fixture in parsed timeline & 1\\
& Read user timeline & HTTP success and stable fixture in parsed timeline & .875\\
\bottomrule
\end{tabularx}
\end{table}

SDK modes are verified from retained application span coverage. The absent mode disables the relevant application tracing providers, samplers or exporters; SDK libraries can remain installed. Partial mode suppresses tracing in the controlled service while retaining the designated surrounding instrumentation. Full mode enables the benchmark's configured native tracing. The experiment therefore evaluates acquisition without adding application SDK instrumentation, with benchmark deployment and tracing-mode adaptations retained as part of preparation. OBI's configured active context propagation remains enabled. Collector releases are OBI 0.15.2, Coroot 1.37.0 and DeepFlow 7.2.3; executable tool commits and image digests are pinned in the artifact.

\subsection{Allocation and independent deployments}
Capture and validation orders are independently randomized and recorded. Petclinic/full/9 and OTel/full/9 provide eight physically paired cases: within each phase, their four treatments run sequentially on the same worker with fresh guest overlays and application state. Capture and validation still use separate workers and deployments. The other 100 cases allocate treatments to separate workers. Logical blocks preserve common fixtures, identities, routing, resource settings, operation semantics and intervention laws; separate workers do not establish physical pairing for overhead or individual runtime-causality claims.

Each sealed blueprint fixes application source, images, SDK settings, routing, resources and initial fixtures. The unintended checkout container-memory restriction identified in earlier preparation was corrected uniformly before this study. Fresh boot and process identities establish deployment provenance without adding graph relations. Execution used standard Ubuntu 24.04 workers with four available CPU cores, approximately 16~GB RAM and a 360-minute job limit. Memory-heavy OTel model processing used its recorded child-process garbage-collection setting and additional owned swap profile; these processing resources are retained separately from application treatment effects.

\subsection{Workload and prospective validation}
Original capture begins with 600 family-preparation attempts and four discovery blocks of 6,000 attempts, covering the controlled replica states. The complete capture is closed before model processing. All four treatment forecasts in a logical block must pass the freeze barrier before independent target deployments begin. Family definitions remain frozen during subsequent evaluation.

The primary target schedule contains eight blocks of 1,200 attempts: four in state $11$, two in $10$, one in $01$ and one in $00$. Operations are balanced within blocks. The nominal offered rate is 20 operation attempts per second, and each case contributes 9,600 primary target attempts under the supplied mixture. The subsequent family test contains 600 attempts. The targets exercise fresh executions of states represented during capture; they do not test transfer to previously unseen intervention types. Setup and readiness traffic and secondary controls are accounted for separately.

No additional experimental response deadline defines primary success. In the OTel replica router, the adapter explicitly sets the Envoy route timeout to \texttt{0s}, disabling that route deadline; removing the injected connection-timeout parameter restores Envoy's default of 5~s~\cite{envoyTimeouts}. Other application limits and transport behaviour remain part of the deployed execution. Workload responses, content checks, SQL postconditions and order-persistence checks supply external evaluation evidence. These labels do not fit the graph or forecast. Pre-freeze verification traffic, when present, remains in the source-global observation prefix. Database-up/down controls retain independent outcomes. OTel persistence controls additionally examine normal execution, a paused consumer, resumption and recovery; persistence measurements alone do not establish an inferred asynchronous causal chain or a model-based delivery forecast. Table~\ref{tab:model-conditions} separates configured conditions, observations and assumptions.

\begin{table}[tbp]
\caption{Status of the conditions used in the structural model and its validation. Configuration records establish the declared setting; they do not certify every runtime condition for every attempt.}
\label{tab:model-conditions}
\centering\small
\begin{tabularx}{\linewidth}{@{}>{\raggedright\arraybackslash}p{23mm}>{\raggedright\arraybackslash}p{44mm}>{\raggedright\arraybackslash}X@{}}
\toprule
Status & Condition or evidence & Scope and retained basis\\
\midrule
Fixed / supplied & Operation contracts, identity groups, replica placement and state law & Pinned declarations and sealed deployment blueprints; these facts are inputs to all methods\\
\addlinespace
Fixed & Controlled-replica schedule and OTel router settings & Registered state blocks; Envoy route deadline disabled by \texttt{0s}, connection default 5~s\\
\addlinespace
Observed & Entry evidence, service identities and dependency edges & Frozen source-global observations and graph annotations; no claim of per-request co-occurrence of every required edge\\
\addlinespace
Measured & Primary success and selected storage postconditions & External responses and content/SQL/order checks under the operation-specific contracts\\
\addlinespace
Assumed fulfilled & Nonintervened requirements, delivery, replica selection, pool and timing behaviour & Fixed guards in the primary structural predicate; complete fulfilment was not independently established for every attempt\\
\addlinespace
Assumed fulfilled & Missing nonintervened database channels & Explicit primary assumptions, excluded from recovered dependencies; selected database-fault queries need channel and fault attribution\\
\addlinespace
Supplied admission rule & Social Network normal path for the ACK model & Conservative requirements for graph support; the list is not an unconditional conjunction defining HTTP acknowledgement in every failure mode\\
\bottomrule
\end{tabularx}
\end{table}

\subsection{Prospective correction of Social Network semantics}
The initial Social Network analysis measured HTTP acknowledgement for \texttt{compose\_post} but imposed an unconditional MongoDB-write requirement. Source inspection established a mismatch between that requirement and the observed event: the pinned Lua client decodes a declared \texttt{ComposePost} service exception without throwing it, so the HTTP handler can return its acknowledgement after the declared error. This possibility does not establish which branch every individual request followed. HTTP acknowledgement is not evidence of a confirmed MongoDB write or durability.

The corrected analysis reuses the exact original observed graphs, entry proofs and model inputs for all 36 Social Network cases. This correction performs no new observation acquisition or Bering discovery; it recompiles the supplied outcome predicates against the frozen graphs. The supplied semantics are applied equally to native, eBPF, combined, context-union and direct methods. Corrected predicates are evaluated by the pinned Sheaft executable and checked against the deterministic graph law. The primary model retains its fail-stop transport and fulfilled-prefix assumptions, with the normal path used as a conservative graph-admission requirement.

For the selected MongoDB-down intervention with replicas up, the corrected compose reference is one under the registered declared-error mask and fixed prefix conditions. The home-timeline reference is one because the selected UserTimeline database is outside that read's path. The user-timeline reference is also one under the preserved warm Redis fixture and fulfilled PostStorage condition: the MongoDB fallback can still be attempted, but the pinned handler ignores its failed cursor and uses retained Redis identifiers. This source, fixture and error information is common to every method. Missing graph structure or selected-database attribution leaves a graph forecast unavailable; it does not suppress the equally informed direct reference.

All nine corrected forecast blocks were sealed before any new target. Fresh independent validation deployments then checked all 36 cases. The first correction series completed measurements but failed to preserve the raw target publications. Its aggregates and costs remain historical evidence. All 36 cases were selected for replacement because of that publication failure, before comparing forecast accuracy, and the replacement used the same frozen forecasts. The final analysis combines the 72 original Petclinic/OTel cases with these 36 corrected Social Network checks. Earlier Social Network results and the first correction aggregates do not add independent replications.

\subsection{Measurement units and interpretation}
For operation $u$ in case $c$, let $S_{cu}$ and $N_{cu}$ be the independently checked successful and total primary attempts. The empirical success frequency is $\widehat q_{cu}=S_{cu}/N_{cu}$. For an available forecast $a_{cum}$ from observation method $m$, the discrepancy is
\begin{equation}
 d_{cum}=\lvert a_{cum}-\widehat q_{cu}\rvert.
 \label{eq:error}
\end{equation}
Support and discrepancy are reported separately. An unavailable model has no numerical forecast error. The support denominator counts case--operation sets and does not weight an application by its request successes. Source/view rows share external outcomes. For each replica state $x$, we also retain its success frequency $S_{cux}/N_{cux}$ and discrepancy from $\psi_u(x)$; the aggregate frequency follows the registered state mixture. These quantities describe different intervention scopes. Every target attempt remains in its original denominator, including failures whose runtime cause is unresolved. Database controls and persistence results retain separate denominators.

Fixed intervention blocks, shared fixtures, repeated requests and only three repetitions per configuration limit statistical generalization. Source projections and graph views cannot be treated as independent observations for binomial confidence intervals. Differences between physical workers and the partially paired allocation also limit causal attribution of treatment effects. A forecast miss remains a result even when its runtime cause is unresolved.

The cost inventory retains preparation, captures, model and baseline processing, simulation, deployment, validation, transfers, publication, failed attempts and technical replacements. Nested phase telemetry is not added again to whole-worker elapsed intervals. Unknown platform intervals and unmeasured human work remain unknown. These intervals are neither CPU time nor billing measurements, and the study does not establish a complete cost advantage over SDK instrumentation or the direct reference.

\paragraph{Generative-AI assistance.}
ChatGPT and Codex assisted with source inspection, extraction and checking of retained records, figure programming, and manuscript drafting and revision. Quantitative findings derive from the identified experimental records; preparation of this manuscript did not generate new application measurements. Automated agent reviews are internal workflow checks and do not constitute external human peer review. The author retains responsibility for the scientific claims and submitted text.

\section{Results}\label{sec:results}

\subsection{Analysis population and model support}
The final population contains 108 cases and 360 distinct case--operation outcome sets. Its 1,036,800 primary attempts comprise 942,168 successes and 94,632 failures, with no unknown or invalid primary outcomes. Each outcome set is reused by the four observation methods and two graph views. The resulting 2,880 comparison rows therefore describe 360 external outcome sets. Primary support and predictions are identical between the raw and network-scoped views, so the primary presentation collapses these duplicate views. Database support differs between views and is reported separately.

\begin{figure}[t]
\centering
\includegraphics[width=\linewidth]{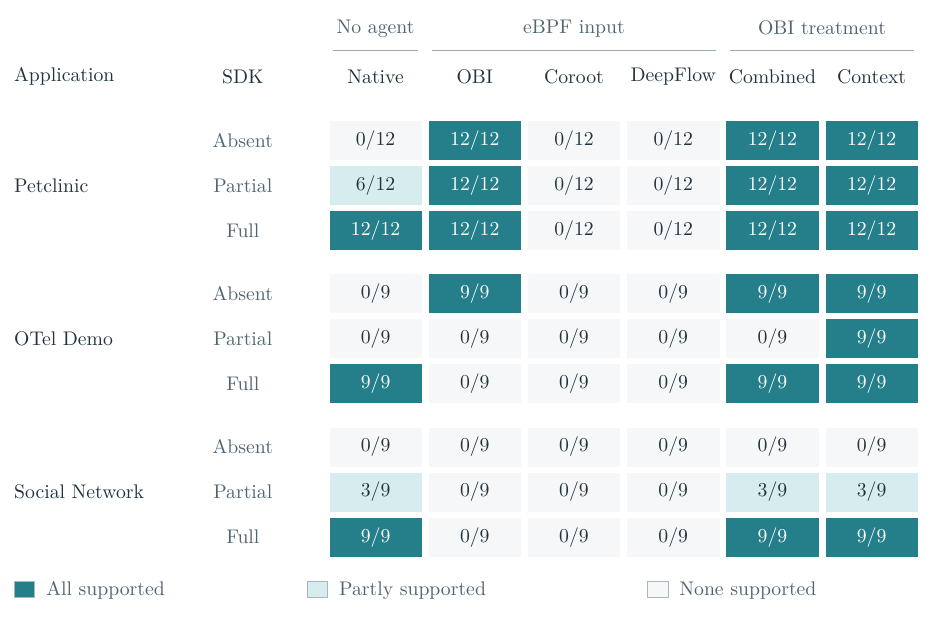}
\caption{Supported primary operation models over three repetitions. Every cell states the number supported and the number attempted. Native control uses the no-agent treatment; the eBPF columns use the corresponding collector treatment; Combined and Context use the OBI treatment. The latter columns therefore compare methods within OBI, rather than additional runtime treatments. Blank evidence produces a recorded zero. Full method-by-treatment results are provided in Supplementary Material.}
\label{fig:coverage}
\end{figure}

Of the 1,440 method comparisons after this collapse, 573 have a supported primary forecast and 867 remain unavailable. At least one method supports a forecast for 192 of the 360 outcome sets. These counts describe the specified acquisition-and-binding procedures; unavailable entries retain their denominators throughout the analysis. Figure~\ref{fig:coverage} reports support for the clean native control, each eBPF-only path, and the two mixed-source methods in the OBI treatment. The supplementary tables cover every treatment and method.

\subsection{Model discovery without SDK tracing (RQ1)}
The tested OBI path supplied primary models for all four Petclinic operations and all three OTel Demo operations with SDK tracing absent. All 21 operation--repetition groups were supported, whereas native input in the no-agent absent-SDK control supported none. This result establishes a working acquisition route in the two tested applications without adding application SDK instrumentation. Its mechanism includes OBI's configured active context propagation, the retained export path and the model-input codec.

The 12 Petclinic groups agreed exactly with their conditional forecasts. OTel Demo contributed five exact agreements and four discrepancies. In repetition 9, browse product produced 2,702 successes in 3,200 attempts, add to cart 2,701, and checkout 2,665, against the forecast .875 for each operation. The respective discrepancies were 3.0625, 3.09375 and 4.21875 percentage points. In repetition 11, browse product produced 2,799 successes in 3,200 attempts, a discrepancy of .03125 percentage points. The remaining absent-SDK OBI groups agreed exactly. Thus 17 of the 21 groups agreed with the supplied conditional abstraction; four exposed additional observed failures. Six exact groups had forecast one; among the fifteen groups with forecast .875, eleven agreed exactly. The mean of the six case-level mean absolute errors was .578125 percentage points. This average is descriptive and does not treat requests or source projections as independent replications.

Figure~\ref{fig:worked-model} illustrates the saved OTel/absent/OBI/repetition-9 checkout model. Its required application subgraph contains the frontend, cart, checkout, currency, payment, shipping, product-catalog and Valkey services. Fourteen required edge records represent nine directed service pairs; the figure collapses parallel records for readability. Product Catalog's two replicas and placement are supplied deployment facts. Its PostgreSQL requirements remain declared fulfilled conditions in the primary query because the selected channel could not be resolved. Consequently the primary forecast is supported while its database-fault forecast remains unavailable.

The saved state responses are $(1,1,1,0)$; the registered mixture gives .875. Subsequent checkout outcomes give $2665/3200=.8328125$. All 135 additional failures occurred in state $10$, with 665 successes in 800 attempts. The state-specific discrepancy is therefore 16.875 percentage points; its weight of $1/4$ gives the mixture discrepancy of 4.21875 percentage points. This localization does not identify the cause. The example demonstrates successful acquisition of the studied model and an observable limit of its conditional abstraction. It was retained as an inspectable example, including the discrepancy, rather than selected to represent every configuration or an error-free result.

The tested Coroot and DeepFlow eBPF-only paths supplied no supported primary operation forecast in this matrix. Their retained observations did not satisfy the declared entry and structural admission requirements. Social Network's absent-SDK cases also remained unsupported with OBI. These results delimit the collector, exporter, codec and consumer configurations exercised here. They do not measure every tracing mechanism available in those projects or establish that no useful model can be obtained from their observations under a different abstraction.

The saved refusal records identify the missing binding evidence more precisely. For Petclinic/absent/repetition~9 list owners, both Coroot and DeepFlow eBPF/raw records resolved the gateway and Customers identities but recorded \texttt{required\_call\_missing\_or\_ambiguous} for that call and \texttt{required\_caller\_not\_uniquely\_reachable} for its MySQL obligation. Coroot additionally recorded \texttt{entry\_endpoint\_missing\_or\_ambiguous} with no candidate for \texttt{GET /api/customer/owners}; DeepFlow's record contained no corresponding entry refusal. OTel/absent/repetition~9 checkout similarly retained resolved frontend and Product Catalog names with an unresolved \texttt{GetProduct} call in both treatments. These are admission failures recorded at the model consumer. The records do not determine which earlier collection, export or correlation stage first lost the required evidence.

\subsection{Partial tracing and source combination (RQ2)}
For partial-SDK Petclinic, the clean native control supported six of twelve operation--repetition groups. OBI eBPF input supported all twelve, as did combined and observed-context input. The missing SDK coverage therefore did not prevent construction of the declared operation models through this eBPF path.

Partial-SDK OTel Demo illustrates the effect of observation-source handling on model support. In the OBI treatment, neither native, eBPF-only nor canonical combined input supported the three operation models. The specified union of observed SERVER-child edges supported all nine operation--repetition groups. Eight forecasts agreed exactly with subsequent outcomes; checkout in repetition 9 produced 2,799 successes in 3,200 attempts. Across the complete study, these nine groups are the only support gains of context union over canonical combination. No group was supported only by combination, and predictions did not differ where both methods were supported.

Inspection of the frozen context sidecars confirmed 17 added SERVER-child edge records in each of the six representations: repetitions 9--11 in the raw and network-scoped views. All original edge records, services and endpoints remained unchanged. These six representations reuse the three captured cases. The added-edge identifiers and their provenance are retained in the evidence files described in Online Resource 1, Section~S9. Exact native/eBPF source--span pairs remain unidentified by this sidecar audit.

The baseline retains the same mixed observations and adds only qualifying observed edges. Under this union, the three partial-SDK OTel operation models became admissible in each repetition. The procedure preserves the scope of the recorded evidence and does not infer missing request ancestry or introduce a new correlation algorithm.

Full SDK coverage did not make every projection interchangeable. The clean native OTel control supported all three operations per repetition. Within the OBI treatment, its native projection supported two, its eBPF projection supported none, and combined input supported all three. Full-SDK acquisition therefore depended on the interaction between the deployed observation treatment and the selected input projection. The experiment did not isolate a single causal mechanism for this difference. Native projections within agent treatments must consequently retain their treatment labels when compared with the clean no-agent control.

\subsection{Forecast agreement across the full matrix}
Every available graph-based forecast equals the equally informed direct conditional reference. The direct calculation needs the supplied operation contract, replica inventory and state law; it does not require an acquired graph. Agreement with this reference verifies the numerical consequence of the stated conditional model, while agreement with subsequent application outcomes tests its adequacy in the executed scenarios.

\begin{figure}[t]
\centering
\includegraphics[width=\linewidth]{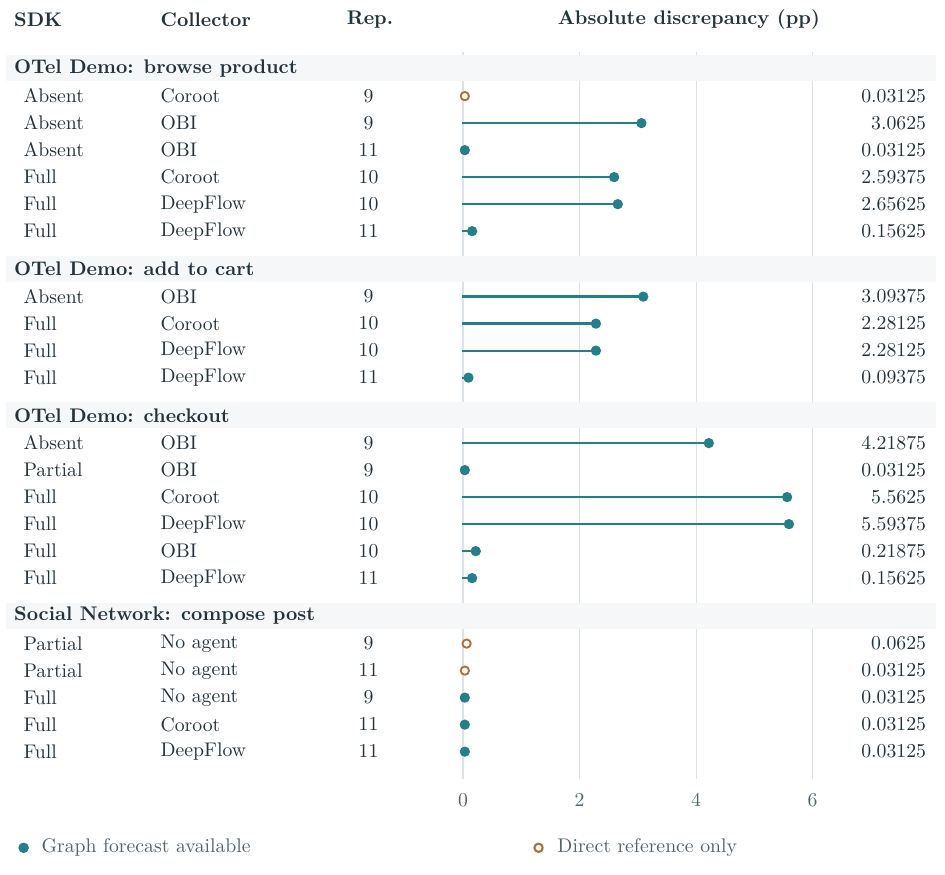}
\caption{All nonzero discrepancies between the direct conditional reference and independent primary outcomes, one point per distinct case--operation set. Filled points have at least one supported graph forecast, which equals the direct reference; open points have no supported graph forecast. The other 339 outcome sets agree exactly with the direct reference, including 174 with a supported graph forecast. Labels give application, SDK mode, runtime treatment, repetition and operation. The axis is in percentage points; printed values preserve the small discrepancies.}
\label{fig:errors}
\end{figure}

Among the 192 distinct outcome sets with at least one available graph forecast, 18 have nonzero discrepancy. The largest is 5.59375 percentage points for OTel/full/DeepFlow/repetition-10 checkout: 2,621 successes in 3,200 attempts against .875. The direct reference differs from the observed outcome in 21 of all 360 sets; three of these have no supported graph forecast. Figure~\ref{fig:errors} shows all 21 discrepancies and distinguishes those two support conditions. The supplementary table preserves exact success counts and denominators. No Petclinic outcome set differs from its applicable direct reference.

The state-specific results expose the size of errors within a single intervention (Table~\ref{tab:state-discrepancies}). The largest is 22.375 percentage points for the same DeepFlow checkout case in state $10$; Coroot/full/repetition~10 checkout has 22.25 percentage points in that state. Across all 1,440 distinct case--operation--state rows, the direct reference has 23 nonzero discrepancies: 15 in $10$, four in $01$ and four in $11$, with none in $00$. Nineteen of these rows have at least one supported graph forecast; four have only the direct reference. The state-specific and mixture discrepancies share the same retained attempts and answer different intervention questions.

\begin{table}[htbp]
\caption{Checkout discrepancies in state $10$ and under the registered mixture. The state forecast is one; the mixture forecast is .875. Other states have no additional failures in these three cases, so multiplying the state discrepancy by $\pi_{10}=1/4$ gives the mixture discrepancy. All errors are in percentage points.}
\label{tab:state-discrepancies}
\centering\small
\begin{tabularx}{\linewidth}{@{}>{\raggedright\arraybackslash}Xrrrr@{}}
\toprule
OTel case & $S_{10}/N_{10}$ & State error & $S/N$ & Mixture error\\
\midrule
Absent / OBI / 9 & 665/800 & 16.875 & 2665/3200 & 4.21875\\
Full / Coroot / 10 & 622/800 & 22.25 & 2622/3200 & 5.5625\\
Full / DeepFlow / 10 & 621/800 & 22.375 & 2621/3200 & 5.59375\\
\bottomrule
\end{tabularx}
\end{table}

The largest discrepancies concern OTel repetitions with additional failures in live-replica states, especially state $10$. The current records do not identify a single cause that explains these failures. The earlier checkout container-memory restriction had already been corrected before V6, so it cannot serve as an established explanation for these results. Small Social Network compose discrepancies remain visible under the corrected acknowledgement contract as well. Model support and empirical accuracy are separate findings: a successfully bound model can still omit conditions that affect operation success.

\subsection{Protocol scope and outcome semantics (RQ3)}
Raw primary support and database-fault support answer different questions. All 249 available database-fault forecasts occur in the network-scoped view: 201 in Petclinic and 48 in corrected Social Network. They cover 87 distinct case--operation outcome sets in 39 cases, with 477 database-down attempts and zero observed discrepancy. Each group contributes five to seven attempts: the 63 Petclinic groups contribute 315 expected failures, and the 24 Social Network groups contribute 162 expected successes. All OTel database-fault forecasts remain unavailable. Duplicate source/method forecasts share the underlying checks, so 249 does not denote 249 independent fault experiments. These small control groups establish agreement in the executed checks; they do not estimate rare-failure probabilities.

The 48 supported Social Network database forecasts occur only in the combined and context methods of the Coroot and DeepFlow treatments. Full SDK supports the three operations per repetition, while partial SDK supports the home-timeline operation. These forecasts cover 24 distinct outcome sets. Native-only and eBPF-only projections did not supply the selected-channel evidence needed for those queries.

The corrected Social Network validation contains 345,600 primary operation attempts. Compose post succeeded in 100,794 of 115,200 attempts, home timeline in all 115,200, and user timeline in 100,800. Under the selected database-down controls, all 720 checked operations succeeded, including 252 compose acknowledgements. This count covers all 36 corrected Social Network cases, including queries without graph support; 162 of those checks belong to the supported subset above. The compose outcomes establish the measured HTTP acknowledgement. They do not establish successful MongoDB writes or durable storage, and the aggregate outcomes do not identify the exception branch taken by every request. The read-user-timeline result is conditional on the retained warm Redis fixture and the pinned fallback handler. The corresponding graph and direct methods use the same corrected information.

OTel's independent persistence controls also separate response acknowledgement from downstream state. The saved checks cover normal processing, a paused consumer, resumption and recovery. These controls test persistence of the identified orders under those conditions. They do not provide a supported asynchronous delivery forecast from the studied structural model. This boundary is consistent with the prior modelling framework, in which delivery, mandatory completion and deadline satisfaction require their own coordinates and evidence~\cite{krasnovsky2026reduced}.

\subsection{Classification, latency and recorded effort}
The supplementary data retain the saved family-classification summaries, unknown-entry counts and case-level latency statistics. Their roles are diagnostic. The saved classification comparisons all report equality with the simple base entry-family baseline. Adjusted Rand index (ARI) scores of one refer to the known assignments; unknown entries retain their denominators. Such agreement does not establish an operation-success predicate or prove that all required channels were recovered. Latency summaries describe the saved Petclinic and OTel per-case request distributions; request arrays are not available for every historical case in the compact publication package. We therefore report no newly pooled latency quantiles and no causal ranking of collector overhead. Eight physically paired cases and 100 between-worker allocations cannot support a uniform paired-overhead interpretation.

The finite cost inventory contains 1,604,380 known worker-seconds, or 445.66 worker-hours. This subtotal combines the original finite metadata cut, its closing intervals and the correction cut without adding nested phase measurements twice. It retains preparation, failed and repeated work. Sixty interval rows in the original cut remain unknown; one correction interval has inconsistent platform timestamps. Successful finalization and later delivery work fall outside the finite subtotal. Human work is unmeasured. Accordingly, this quantity is an incomplete record of elapsed worker allocation, not CPU time, billing, sequential elapsed time or a measured saving over introducing SDK tracing.

\section{Modelling implications}\label{sec:discussion}

\subsection{Extending model discovery to underinstrumented applications}
The OBI results extend the demonstrated acquisition conditions of the model-discovery and graph-simulation approach. In the tested Petclinic and OTel deployments, observations supported the operation models in all 21 absent-SDK operation--repetition groups. The resulting representations were executable in the existing toolchain and supplied forecasts for the declared replica-failure scenarios. This closes the acquisition step for those cases even though application SDK tracing was disabled.

For a compatible operational legacy application, the practical opportunity is to begin virtual fault experiments on the current deployment before completing a tracing retrofit. Migration and refactoring can follow their own requirements and schedules. The modelling workflow still requires collector deployment, identity resolution, operation semantics and explicit scenario assumptions. The experiment established feasibility under controlled tracing deficits; transferring it to an industrial installation requires checking its actual platform, protocols and available declarations.

The scope of the intended forecast determines the evidence required. This experiment investigated conditional structural availability under a supplied replica-state law. Response-time prediction, load-dependent failures, durable completion and previously unseen recovery behaviour require further model coordinates and observations. The role of the present acquisition result is to make the declared fault analysis available under limited tracing, with its assumptions and unsupported cases retained.

For this instance, the equally informed direct calculation and graph-based models evaluate the same supplied replica condition. Discovery produces a dependency representation with inspectable provenance that can be resolved against operation requirements and executed by the simulator. Equality of numerical forecasts provides limited evidence about structural accuracy because the primary predictions take only two values. The retained source-level witnesses and replay in Online Resource~2 therefore document the acquired model separately from forecast agreement.

\subsection{Model adequacy requires an explicit success event}
The Social Network correction demonstrates why operation semantics belong in the model specification. HTTP acknowledgement and storage completion are distinct observable events. The preceding unconditional write contract imposed a stronger requirement than the measured compose outcome. Recompiling a corrected contract against the same graph, then validating it on fresh targets, aligned the model with the registered event without attributing new edges to the observation mechanism.

The correction also limits the interpretation of unsupported cases. Admission uses the requirements of a particular model. An acquisition path that fails those requirements has not established an executable instance of that model; the failure does not prove that every possible model of the application's acknowledgement is unattainable. Conversely, a supported binding does not establish that the selected abstraction includes every cause of failure. The OTel discrepancies test this distinction directly: the worked example differs by 16.875 percentage points in state $10$, and the full matrix reaches 22.375 percentage points in that state. The registered mixture reduces their aggregate contribution, while leaving the state-specific limitation unchanged. The retained observations do not isolate its cause.

The formal model in Section~\ref{sec:model} gives a route for specifying richer obligations. Delivery to a replica, completion of mandatory calls and satisfaction of a deadline can be represented separately when suitable evidence is available. In the current implementation, fixed fulfilled conditions and structural reachability make those boundaries explicit. Preserving unsupported queries and empirical discrepancies prevents graph connectivity from being interpreted as evidence of storage durability or end-to-end completion.

\subsection{Evidence association and deployable observation paths}
The partial-SDK OTel results show that the source view used by a model consumer can affect model support even when the underlying observations are fixed. The observed-context union recovered nine previously unsupported case--operation groups using existing endpoints and qualifying observed parent--child edges. This finding motivates examination of evidence association at collector/exporter boundaries before introducing a more elaborate reconstruction algorithm.

Existing eBPF systems address a broader range of correlation and protocol mechanisms than the exported paths exercised here~\cite{shen2023,geng2025deeptrace,geng2026flowscope,hong2026}. The unit of comparison in this study is therefore the configured observation-to-model path. Extending a codec or adopting a different correlation method could change its support boundary. Any such extension would need its own provenance rules and a fresh prospective check, especially if it inferred dependencies rather than preserving observed relationships.

\section{Threats to validity}\label{sec:threats}

\paragraph{Construct validity.}
The success checks are operation-specific. Petclinic includes payload and SQL postconditions, OTel primary checks stop at the defined synchronous order response, and corrected Social Network compose targets HTTP acknowledgement. These events must remain attached to their results. The state mixture is supplied; the experiment does not estimate operational availability under a production failure distribution. Source-global discovery includes preparation and background traffic, so graph edges do not establish request-specific causality. A qualifying edge observed only in another context can satisfy the implemented structural admission rule. No independent operation-specific graph reference establishes edge precision or recall in this study. Service identities, replica placement, operation semantics and fixed nonintervened conditions remain supplied inputs.

\paragraph{Internal validity.}
Frozen inputs and forecasts precede fresh target deployments, including the corrected Social Network series. Targets nevertheless exercise intervention states already represented during capture. Only eight cases were physically paired within phases; worker and runtime differences can affect the remaining comparisons. OBI's active propagation forms part of the deployed treatment. The data do not isolate the cause of all forecast discrepancies. Saved latency summaries combine operations and failure states; they do not isolate the overhead of SDK tracing or an eBPF collector. Computational costs of the preceding modelling procedure were evaluated separately~\cite{krasnovsky2026reduced}; those measurements exclude telemetry acquisition and are not performance results of the present experiment. Technical replacement of the first corrected series followed loss of its raw publications, with all cases selected before comparing accuracy; failed work and its costs remain recorded.

\paragraph{Conclusion validity.}
Each configuration has three repetitions. Fixed state blocks, common fixtures and repeated operation attempts create dependence. Methods and graph views reuse the same outcomes. Reported counts and discrepancies are descriptive; no claim relies on treating over a million attempts as independent draws from a production population. The 18 graph-supported mixture discrepancies remain separate from three direct-reference discrepancies without graph support; at state level, the corresponding counts are 19 and four. Supported database-fault results use only five to seven attempts per group. The finite cost inventory contains unknown intervals and excludes unmeasured human work, so it cannot establish comparative total effort.

\paragraph{External validity.}
The study uses three public applications, ten operations, the pinned collector releases and a two-guest Linux deployment. These conditions cover useful protocol and tracing contrasts but do not represent all legacy technologies, encrypted traffic, deployment topologies or failure mechanisms. The industrial literature in Section~\ref{sec:legacy-problem} establishes operational motivation; the benchmark experiment establishes feasibility within the evaluated conditions. No production-team adoption study or comparison of developer effort was conducted. In particular, the cited industrial legacy cases are evidence of continued operation and change constraints, with no claim that the tested Linux eBPF path supports their platforms.

\paragraph{Reproducibility scope.}
The compact deposit contains code, pinned specifications and analysis records sufficient for the reported summary reaggregation, together with one worked observation-to-model example. It does not retain every historical raw event, per-attempt latency or classification assignment. The workflow supports new execution of the experimental protocol subject to the documented software and worker requirements; the deposit does not claim byte-for-byte regeneration of every historical trace from the compact records. The selected example illustrates an inspectable acquisition path and is not a representative statistical sample. Supplementary Material identifies this distinction, the selected cohorts and their hashes.

\section{Conclusion}\label{sec:conclusion}
This study extended the model-discovery and graph-simulation approach to virtual fault experiments under absent or incomplete application SDK tracing. Existing eBPF collectors supplied the observation mechanism; the empirical question concerned obtaining an executable operation model and checking its forecasts. In the tested OBI deployments, all seven Petclinic and OTel operations were supported across three repetitions without SDK tracing. Seventeen of the 21 operation--repetition groups agreed exactly with their forecasts; four exposed additional observed failures. The maximum mixture discrepancy was 4.22 percentage points and the maximum state-specific discrepancy in that subset was 16.875 percentage points. Combining observed context links supplied nine additional groups under partial tracing.

The complete 108-case study identifies a usable acquisition path and its boundaries. Protocol support and evidence association determined whether the required representation could be constructed. Operation contracts determined the meaning of the forecast, including the distinction between Social Network HTTP acknowledgement and a successful database write. All unsupported cases and observed discrepancies remain part of the result.

Operational legacy systems motivate this extension because necessary services can remain in use while their components are difficult to instrument or replace. The benchmark evidence supports beginning the declared model-based fault analysis before complete tracing, subject to platform compatibility and the model's semantic requirements. The work leaves the measurement of industrial adoption costs and broader failure behaviour to subsequent studies.

Code, pinned experimental specifications and analytical records, together with one worked observation-to-model example, are archived at Zenodo (DOI \href{https://doi.org/10.5281/zenodo.24131735}{10.5281/zenodo.24131735}). Supplementary Material (Online Resource~1) follows the references; the analytical archive (Online Resource~2, \texttt{ESM\_2.zip}) accompanies this preprint as an ancillary file on its arXiv record.

\backmatter

\bibliography{references}
\clearpage
\begingroup
\backmatterfalse
\setcounter{secnumdepth}{3}
\setcounter{section}{0}
\setcounter{subsection}{0}
\setcounter{subsubsection}{0}
\setcounter{table}{0}
\setcounter{figure}{0}
\setcounter{equation}{0}
\renewcommand{\thesection}{S\arabic{section}}
\renewcommand{\thetable}{S\arabic{table}}
\renewcommand{\thefigure}{S\arabic{figure}}
\renewcommand{\theequation}{S\arabic{equation}}
\renewcommand{\theHsection}{supp.\arabic{section}}
\renewcommand{\theHsubsection}{supp.\arabic{section}.\arabic{subsection}}
\renewcommand{\theHsubsubsection}{supp.\arabic{section}.\arabic{subsection}.\arabic{subsubsection}}
\renewcommand{\theHtable}{supp.\arabic{table}}
\renewcommand{\theHfigure}{supp.\arabic{figure}}
\renewcommand{\theHequation}{supp.\arabic{equation}}
\renewcommand{\baselinestretch}{1}
\makeatletter
\renewcommand\normalsize{\@setfontsize\normalsize\@xipt{14.96pt}%
  \abovedisplayskip 11pt plus3pt minus6pt%
  \abovedisplayshortskip 0pt plus3pt%
  \belowdisplayshortskip 6.5pt plus3.5pt minus3pt%
  \belowdisplayskip\abovedisplayskip\let\@listi\@listI}
\renewcommand\small{\@setfontsize\small\@xpt{13.2pt}}
\renewcommand\footnotesize{\@setfontsize\footnotesize\@ixpt{12.1pt}}
\makeatother
\normalsize
\setlength{\parindent}{0pt}
\setlength{\parskip}{6pt}
\setlength{\emergencystretch}{2em}
\setlength{\tabcolsep}{4pt}
\renewcommand{\arraystretch}{1.16}
\setlength{\LTcapwidth}{\textwidth}
\setlength{\LTpre}{6pt}
\setlength{\LTpost}{6pt}
\newcolumntype{L}[1]{>{\raggedright\arraybackslash}p{#1}}
\newcolumntype{R}[1]{>{\raggedleft\arraybackslash}p{#1}}
\newcolumntype{C}[1]{>{\centering\arraybackslash}p{#1}}
\providecommand{\hash}[1]{\path{#1}}
\raggedbottom
\section*{Supplementary Material}
\addcontentsline{toc}{section}{Supplementary Material}
This appendix contains the Supplementary Material (Online Resource~1). The analytical archive (Online Resource~2, \texttt{ESM\_2.zip}) accompanies the preprint as an ancillary file on its arXiv record.

\section{Study population and analytical records}
This supplement documents the completed V6 experiment and the prospectively evaluated Social Network outcome correction. The manuscript uses these records to evaluate model acquisition for virtual fault experiments under incomplete application tracing. Real application interventions provide the subsequent validation of model forecasts. The reported analysis contains 108 cases: 72 original Petclinic and OpenTelemetry Demo cases and 36 corrected Social Network cases. Each case is identified by application, SDK-coverage mode, repetition and runtime treatment. The three SDK modes are absent, partial and full; repetition identifiers are 9, 10 and 11; runtime treatments are no study eBPF agent (NONE), OBI, Coroot and DeepFlow. A logical block fixes application, SDK mode and repetition and contains the four treatments.

The 360 distinct case--operation units contain 1,036,800 primary attempts: 942,168 successes and 94,632 failures, with no unknown or invalid primary outcomes. All declared operations, unavailable forecasts and adverse outcomes remain in the analysis. Four observation methods and two graph views reuse each case's independently measured target outcomes. Historical Social Network results and the first correction series, whose raw publication failed, remain provenance records. They do not increase the number of independent cases in the final analysis. The accompanying Online Resource~2 (\path{ESM_2.zip}) reorganizes saved analytical values and identifies the selected cases; it does not contain newly collected application observations. Its \path{selected-example} directory provides illustrative model, predicate and forecast records at their original paths, together with a receipt. Execution from the archived observations requires the full selected-case package in the DOI deposit. The additional verification records in Section~\ref{sec:supp-replay} document a completed replay of that package.

The compact artifact identified above contains the complete source snapshot, experiment configurations, original analytical records, the corrected Social Network summary, provenance and cost information, and one selected real capture/model example. The example is OTel Demo with SDK tracing absent, repetition 9, and OBI. It was selected as a small case with usable observation and model packages, and is not a representative statistical subsample. Its model availability does not imply exact agreement of every target outcome with the forecast.

\clearpage
\section{Pinned software and execution configuration}
The following revisions identify the implementation used by the study. They specify executable inputs rather than the capabilities of current product releases. Per-case provenance records identify the actual producing source and image inventories, including technical processing continuations.

\begingroup\small
\begin{longtable}{@{}L{31mm}L{125mm}@{}}
\caption{Application and modelling-tool source pins.}\label{tab:supp-source-pins}\\
\toprule Component & Source revision\\\midrule
\endfirsthead
\toprule Component & Source revision\\\midrule
\endhead
Procrustes & \hash{9648ae3fe9ef24725e665d008c79d0681ce40f85}\\
Bering & \hash{66ca3af1fe581839e525fb9cc5600a1edd6ca8fb}\\
Sheaft & \hash{686a602f4e7b71f3612f1810885e29dd66c8733a}\\
Petclinic Microservices & \hash{3858f9c630cf989bb6809a86edf47c2be78dc9f1}\\
OTel Demo 3.0.0 & \hash{8c47d47c9ac27710d2b2a153bcd53e483bffe66d}\\
DeathStarBench Social Network & \hash{6ecb09706140f8730b5385c08f1386c654c3c526}\\
Coroot 1.37.0 & \hash{89a858a0ee35c4eb938b86d56f23168fb0ac7573}\\
DeepFlow 7.2.3 & \hash{9a1219fa5cf1471ba2ec8742019661567d98002e}\\
\bottomrule
\end{longtable}
\endgroup

\begingroup\small
\begin{longtable}{@{}L{31mm}L{125mm}@{}}
\caption{Selected immutable collector, runtime and infrastructure inputs. SHA-256 values refer to the identified image or file.}\label{tab:supp-image-pins}\\
\toprule Component & Image/file and SHA-256\\\midrule
\endfirsthead
\toprule Component & Image/file and SHA-256\\\midrule
\endhead
OBI 0.15.2 & \path{otel/ebpf-instrument:v0.15.2}\par
\hash{101b5c773ac84545cac007ce3dbaf38bfb0e7091f26e813c08d649f35dca253e}\\
Coroot 1.37.0 & \path{ghcr.io/coroot/coroot-node-agent:1.37.0}\par
\hash{f360c31c5e2fa75550c0029f3c02a0e54b83831bae59201b9fe76bab2fe564b4}\\
DeepFlow 7.2.3 & \path{deepflowce/deepflow-agent:v7.2.3}\par
\hash{5f72c0996fdb5008fdb8fd0524f427a895fc54c61690b26ba2cae648929ea95a}\\
Guest base image & Ubuntu Noble cloud image dated 17 October 2026\par
\hash{04e8f2461a16a44e78a370284471f29616cdd9baf1893df25807b71f985b6c94}\\
Petclinic Java runtime & \path{eclipse-temurin:17-jre}\par
\hash{13cc28a6cc72a38ce1f00c906be3580c1a3e604b8984d694f369a96742abc93b}\\
MySQL 8.4 & \path{mysql:8.4}\par
\hash{b3b90af2a6552ae30c266fdb7d5dd55f3afb72404bb78d37fe8a23eb857fd3fb}\\
HAProxy 3.0 & \path{haproxy:3.0-alpine}\par
\hash{34cc7d1f6142464d7d2b73e2a1eef7392556dbf304160aef543e513cfd9e5162}\\
OTel Java agent 2.31.1 & \path{opentelemetry-javaagent.jar}\par
\hash{bbf83c151b6400709e2f225bdd07a04f839d9d13b8b93464241333fd25d3e3ba}\\
OTel Collector Contrib 0.159.0 & \path{ghcr.io/open-telemetry/opentelemetry-collector-releases/opentelemetry-collector-contrib:0.159.0}\par
\hash{1f2c54a30e713fac6b3ae77a1ec84010c2007e29ced8ec666214fc2f6739c1cc}\\
Social Network service image & \path{deathstarbench/social-network-microservices:latest}\par
\hash{a845266e98ce15d204ee7b98d41a2a95820af525013ce70f14104f5b9b5e92b6}\\
Social Network MongoDB & \path{mongo:4.4.6}\par
\hash{6efa052039903e731e4a5550c68a13c4869ddc93742c716332883fd9c77eb79b}\\
\bottomrule
\end{longtable}
\endgroup

The worker label was \path{ubuntu-24.04}, with four available CPU cores, approximately 16~GB RAM and a 360-minute job limit. A deployment used two QEMU/KVM guests with separate kernels, each with two virtual CPUs. The provisioner requested 6,144~MiB per guest and used its recorded 3,072~MiB fallback when host memory was below 14,000,000~KiB. The sealed blueprint and per-phase resource receipts identify the actual allocation. Validation created fresh guest overlays and application state from that blueprint; retained boot and process identities distinguish the deployments.

Petclinic/full/9 and OTel/full/9 comprise eight physically paired cases. Within each phase, their treatments ran sequentially on one worker with fresh guests and state; capture and validation still used separate workers. The other 100 cases used separate treatment workers. Capture and validation orders were independently randomized and retained. These logical blocks hold fixtures and settings common while preserving actual host differences; the separately allocated cases do not supply physical pairing for an overhead comparison.

Memory-heavy OTel model processing used the recorded Bering-child \path{GOGC=25} setting and an additional owned 16~GiB swap file. These model-processing settings are distinct from collector effects during application execution. The unintended checkout container-memory limit identified in earlier preparation was corrected uniformly before V6; it is not an established explanation of discrepancies retained in the current results.

\subsection{Tracing modes and evidence conversion}
Petclinic absent mode omitted the Java agent and disabled application SDK export; partial mode retained instrumentation on Gateway, Customers and Vets while disabling both Visits replicas. OTel absent mode applied the recorded provider/exporter/sampler guards to application services; partial mode disabled both Product Catalog replicas. Libraries or propagation code could remain installed when recording/export was disabled. Social Network changed the C++ and nginx Jaeger configurations: absent used the disabled setting and constant-zero sampling; partial disabled both User Timeline replicas; full used the configured native tracing. Compliance was checked from actual application spans, not inferred solely from environment variables.

OBI used active context propagation with its Java/Node runtime auto-instrumentation features disabled. Coroot retained its pinned node-agent exports. DeepFlow retained standalone JSONL and converted supported records into the model-discovery input without inventing trace, span, parent, request, message or service identities. The conversion admitted unambiguous eBPF records for HTTP, HTTP/2, gRPC, MySQL, PostgreSQL, Redis, MongoDB, Memcached and Kafka. No specialized framed binary Thrift decoder was integrated. The experiment therefore addresses these complete configured acquisition paths.

\section{Operation contracts and independent outcome checks}
The workload supplied opaque request identifiers and attempt/step boundaries. Entry matching could use an observed identifier or an admitted service/method/route match. Discovery consumed the complete closed source prefix, including preparation, verification and background/control traffic. Operation requirements, identity aliases and replica placement were supplied equally to all methods. They did not insert missing dependency edges.

The controlled services were Visits, Product Catalog and User Timeline, with one replica per guest. Each original capture retained 600 family-preparation attempts followed by four discovery blocks of 6,000 attempts. Primary validation used eight blocks of 1,200 attempts, with replica-state multiplicities 4/2/1/1 for $11/10/01/00$ and a nominal rate of 20 operation attempts/s. The family test contained 600 attempts. Model inputs were closed and all treatment forecasts frozen before target deployments. The Social Network correction reused the original graphs and family definitions and acquired new validation observations.

\begingroup\small
\begin{longtable}{@{}L{36mm}L{120mm}@{}}
\caption{Primary external outcomes and graph-admission obligations for all ten operations. HTTP steps within an operation do not count as independent experimental repetitions.}\label{tab:supp-contracts}\\
\toprule Application and operation & Contract and observation boundary\\\midrule
\endfirsthead
\toprule Application and operation & Contract and observation boundary\\\midrule
\endhead
Petclinic\par\path{list_owners} &
\textbf{Request:} GET \path{/api/customer/owners}.\par
\textbf{Success:} HTTP 200 and canonical owner-list JSON matching the SQL-derived fixture.\par
\textbf{Admission:} Gateway--Customers and the declared Customers--MySQL read channel. Nonintervened database fulfilment is an explicit primary condition.\\
\addlinespace
Petclinic\par\path{list_vets} &
\textbf{Request:} GET \path{/api/vet/vets}.\par
\textbf{Success:} HTTP 200 and canonical veterinarian payload matching the fixture.\par
\textbf{Admission:} Gateway--Vets. The MySQL requirement is conditional on cache miss; missing evidence for that condition remains unknown for database-fault analysis.\\
\addlinespace
Petclinic\par\path{owner_details_with_visits} &
\textbf{Request:} GET \path{/api/gateway/owners/6}.\par
\textbf{Success:} HTTP 200 with the complete expected owner, pets and visits. An incomplete fallback payload fails.\par
\textbf{Admission:} Gateway--Customers and Gateway--Visits, with declared Customers/Visits database reads.\\
\addlinespace
Petclinic\par\path{create_visit} &
\textbf{Request:} POST \path{/api/visit/owners/1/pets/1/visits}.\par
\textbf{Success:} HTTP 201, the expected returned marker, and an independent exact SQL-row match for the unique marker, returned identifier, pet, date and description. Unavailable SQL evidence remains unknown.\par
\textbf{Admission:} Gateway--Visits and the declared Visits--MySQL write. The SQL oracle is evaluation evidence, not discovery input.\\
\addlinespace
OTel Demo\par\path{browse_product} &
\textbf{Request:} GET \path{/api/products/OLJCESPC7Z}.\par
\textbf{Success:} HTTP 200 and matching product identifier.\par
\textbf{Admission:} Frontend--Catalog \path{GetProduct} and the declared Catalog--PostgreSQL read.\\
\addlinespace
OTel Demo\par\path{add_to_cart} &
\textbf{Requests:} product GET followed by POST \path{/api/cart}.\par
\textbf{Success:} Both steps succeed, the product matches, and the selected cart item has quantity one.\par
\textbf{Admission:} Browse obligations; Frontend--Cart \path{AddItem}/\path{GetCart}; declared Cart--Redis write/read.\\
\addlinespace
OTel Demo\par\path{checkout} &
\textbf{Requests:} product GET, cart POST and POST \path{/api/checkout?currencyCode=USD}.\par
\textbf{Success:} All required HTTP steps succeed; product/cart checks pass; the response contains an order identifier and the matching item at quantity one.\par
\textbf{Admission:} Add-to-cart obligations; Frontend--Checkout; Checkout calls to Cart, Catalog, Currency, Shipping and Payment; and Frontend--Catalog response enrichment, with the corresponding declared database channels. Asynchronous persistence is a separate outcome.\\
\addlinespace
Social Network\par\path{compose_post} &
\textbf{Request:} POST \path{/wrk2-api/post/compose}.\par
\textbf{Success:} HTTP success and the exact \emph{Successfully upload post} acknowledgement. This outcome does not assert a MongoDB write or durability.\par
\textbf{Primary admission:} The conservative normal-path list covers Nginx--Compose; Compose calls to Text, User, Media, UniqueID, PostStorage, UserTimeline and HomeTimeline; and declared storage/social-graph obligations. This list is not an unconditional conjunction defining ACK under every failure mode. Corrected database-down semantics are specified below.\\
\addlinespace
Social Network\par\path{read_home_timeline} &
\textbf{Request:} GET \path{/wrk2-api/home-timeline/read?user_id=3&start=0&stop=10}.\par
\textbf{Success:} HTTP success and a valid parsed timeline containing the stable follower fixture.\par
\textbf{Admission:} Nginx--HomeTimeline; HomeTimeline--Redis/PostStorage. The selected UserTimeline MongoDB is outside this read's path.\\
\addlinespace
Social Network\par\path{read_user_timeline} &
\textbf{Request:} GET \path{/wrk2-api/user-timeline/read?user_id=2&start=0&stop=10}.\par
\textbf{Success:} HTTP success and a valid parsed timeline containing the stable owner fixture.\par
\textbf{Admission:} Nginx--UserTimeline and UserTimeline--Redis/PostStorage. MongoDB can be attempted on fallback. The corrected database-down contract uses the registered preserved warm Redis fixture and fulfilled PostStorage condition.\\
\bottomrule
\end{longtable}
\endgroup

The executed V4 adapters used by V6 impose no additional experimental response deadline. The OTel replica router sets the Envoy route timeout to \texttt{0s}, disabling that route deadline; removing the injected connection-timeout parameter restores the default of 5~s. Other application and transport limits remain in effect. Historical adapter documentation also describes earlier deadline policies; those policies do not define the present primary outcomes. Petclinic SQL verification, immediate OTel checkout success, Social Network acknowledgement and asynchronous order persistence have different evidence boundaries and are reported separately.

\subsection{Corrected Social Network database-down semantics}
The selected fault stops the UserTimeline MongoDB container with both controlled service replicas up. The pinned ComposePost processor can return a declared service exception in a normal Thrift reply. Its Lua receiver decodes that exception field without throwing it, after which the HTTP handler can emit the acknowledgement. Under the registered exception mask and fixed prefix conditions, the direct conditional ACK forecast is one. The selected MongoDB remains on the attempted path; it is not reclassified as off-path. Source inspection establishes the possibility of this masking branch, not proof that every observed acknowledgement traversed it.

For \path{read_home_timeline}, the direct database-down forecast is one under the fixed unrelated-service conditions. For \path{read_user_timeline}, successful readiness/reset preserves the untouched fixture in Redis. The pinned handler can retain those identifiers when the MongoDB fallback cursor fails and retrieve the posts from PostStorage. Its direct forecast is one conditional on that fixture, Redis/PostStorage and noncongested pools. Preparation retries can create repeated fixture records; the contract does not require an invented exactly-one-post precondition.

This source, fixture and error information is available equally to all observation methods and the direct comparator. A graph-based database forecast additionally requires admitted structure and attribution of the selected database endpoint; failure to establish those observations leaves it unavailable. The same restriction is not imposed on the equally informed direct calculation, which does not recover a graph.

All nine corrected forecast blocks were frozen before fresh validation. The first correction run completed the 36 measurements but lost the raw target publication. The replacement selected all 36 cases because of that publication failure and reused the exact frozen forecasts. Historical measurements and all associated costs remain retained; neither earlier series adds independent cases to the final analysis.

\subsection{Asynchronous persistence and classification}
OTel controls record independently observed order/item/shipping state during normal execution, a paused accounting consumer, resumption and recovery. In the executed adapter, verification takes a SQL snapshot and preserves exact persistence, mismatch or pending states; it does not impose the earlier adapter's completion deadline. A pending snapshot does not prove eventual loss. Persisted records do not establish that the observation pipeline reconstructed the producer--consumer causal chain, and they are not themselves a model-based delivery forecast.

Operation-family fitting uses frozen method/route/step/source signatures and observed call-set descriptors. The workload's operation labels and external outcome labels do not fit the families. Classification agreement is descriptive and does not replace outcome validation. The compact analytical JSON projections omit \path{frozen_classification.class_test_assignments} and \path{frozen_classification.target_descriptive_assignments}. They also omit individual request durations. Consequently, the artifact supports reaggregation of saved classification and latency summaries but does not support rederiving every historical adjusted Rand index or latency quantile from individual attempts.

\section{Comparison units and complete model coverage}
The case key is application, SDK mode, repetition and runtime treatment. The primary outcome unit adds the operation. A method/view result additionally identifies the observation method and graph view. The method labels are E (eBPF), N (native SDK), C (ordinary combined input) and U (observed-context union). Native observations in the NONE treatment provide the clean SDK control; native observations in an agent treatment remain that treatment's source projection.

The primary support statuses and point predictions agree between raw and network-scope views. The 2,880 method/view rows therefore reduce to 1,440 primary method rows when the duplicate view is removed; 573 of these rows have an available primary forecast and 867 remain unavailable. At the distinct case--operation level, at least one observation method supplies a forecast for 192 of 360 units. Tables that collapse these duplicate primary views count each method--case--operation result once. Database-fault forecasts are view-dependent and retain the original view identifier. All 249 available database-fault rows occur in the network-scope view (201 Petclinic and 48 Social Network), covering 87 distinct case--operation units in 39 cases and 477 database-down attempts. Petclinic contributes 63 units in 27 cases and 315 expected failures; Social Network contributes 24 units in 12 cases and 162 expected successes. Each supported unit contains five to seven attempts. The 720 Social Network database-down attempts in the complete corrected 36-case population form a broader denominator and must not be substituted for these 162 attempts. Every raw-view database forecast remains unavailable. The available database-fault forecasts agree with the equally informed direct reference and the measured outcomes, conditional on the corresponding outcome contract. The CSV records preserve unavailable forecasts separately from numerical discrepancies. They must not be interpreted as zero-error predictions.

The coverage table reports the complete treatment/application/SDK matrix. Its support denominators count operation--repetition groups under the stated method and treatment, not successful requests. The fixed state blocks and repeated source/view analyses do not supply independent trials for confidence intervals. All comparisons remain conditional on the supplied contracts, identities, replica inventory and intervention law.

\begingroup\small
\begin{longtable}{@{}L{32mm}L{17mm}L{26mm}R{17mm}R{17mm}R{17mm}R{17mm}@{}}
\caption{Complete primary forecast availability. Entries are available/declared operation--repetition units for one application, SDK mode, treatment and observation method. Each denominator includes three repetitions of four Petclinic operations or three OTel/Social Network operations. N: native; E: eBPF; C: ordinary combined; U: observed-context union. Raw and network-scope primary views coincide and are counted once.}\label{tab:supp-coverage}\\
\toprule Application & SDK & Treatment & N & E & C & U\\\midrule
\endfirsthead
\multicolumn{7}{@{}l}{\textit{Table~\thetable{} continued}}\\[3pt]
\toprule Application & SDK & Treatment & N & E & C & U\\\midrule
\endhead
\midrule\multicolumn{7}{r@{}}{\textit{Continued on the next page}}\\
\endfoot
\bottomrule
\endlastfoot
Petclinic & absent & NONE & 0/12 & 0/12 & 0/12 & 0/12 \\*
Petclinic & absent & OBI & 0/12 & 12/12 & 12/12 & 12/12 \\*
Petclinic & absent & Coroot & 0/12 & 0/12 & 0/12 & 0/12 \\*
Petclinic & absent & DeepFlow & 0/12 & 0/12 & 0/12 & 0/12 \\
Petclinic & partial & NONE & 6/12 & 0/12 & 6/12 & 6/12 \\*
Petclinic & partial & OBI & 6/12 & 12/12 & 12/12 & 12/12 \\*
Petclinic & partial & Coroot & 6/12 & 0/12 & 6/12 & 6/12 \\*
Petclinic & partial & DeepFlow & 6/12 & 0/12 & 6/12 & 6/12 \\
Petclinic & full & NONE & 12/12 & 0/12 & 12/12 & 12/12 \\*
Petclinic & full & OBI & 12/12 & 12/12 & 12/12 & 12/12 \\*
Petclinic & full & Coroot & 12/12 & 0/12 & 12/12 & 12/12 \\*
Petclinic & full & DeepFlow & 12/12 & 0/12 & 12/12 & 12/12 \\
OTel Demo & absent & NONE & 0/9 & 0/9 & 0/9 & 0/9 \\*
OTel Demo & absent & OBI & 0/9 & 9/9 & 9/9 & 9/9 \\*
OTel Demo & absent & Coroot & 0/9 & 0/9 & 0/9 & 0/9 \\*
OTel Demo & absent & DeepFlow & 0/9 & 0/9 & 0/9 & 0/9 \\
OTel Demo & partial & NONE & 0/9 & 0/9 & 0/9 & 0/9 \\*
OTel Demo & partial & OBI & 0/9 & 0/9 & 0/9 & 9/9 \\*
OTel Demo & partial & Coroot & 0/9 & 0/9 & 0/9 & 0/9 \\*
OTel Demo & partial & DeepFlow & 0/9 & 0/9 & 0/9 & 0/9 \\
OTel Demo & full & NONE & 9/9 & 0/9 & 9/9 & 9/9 \\*
OTel Demo & full & OBI & 6/9 & 0/9 & 9/9 & 9/9 \\*
OTel Demo & full & Coroot & 9/9 & 0/9 & 9/9 & 9/9 \\*
OTel Demo & full & DeepFlow & 9/9 & 0/9 & 9/9 & 9/9 \\
Social Network & absent & NONE & 0/9 & 0/9 & 0/9 & 0/9 \\*
Social Network & absent & OBI & 0/9 & 0/9 & 0/9 & 0/9 \\*
Social Network & absent & Coroot & 0/9 & 0/9 & 0/9 & 0/9 \\*
Social Network & absent & DeepFlow & 0/9 & 0/9 & 0/9 & 0/9 \\
Social Network & partial & NONE & 3/9 & 0/9 & 3/9 & 3/9 \\*
Social Network & partial & OBI & 3/9 & 0/9 & 3/9 & 3/9 \\*
Social Network & partial & Coroot & 3/9 & 0/9 & 3/9 & 3/9 \\*
Social Network & partial & DeepFlow & 3/9 & 0/9 & 3/9 & 3/9 \\
Social Network & full & NONE & 9/9 & 0/9 & 9/9 & 9/9 \\*
Social Network & full & OBI & 9/9 & 0/9 & 9/9 & 9/9 \\*
Social Network & full & Coroot & 9/9 & 0/9 & 9/9 & 9/9 \\*
Social Network & full & DeepFlow & 9/9 & 0/9 & 9/9 & 9/9 \\
\end{longtable}

\endgroup

\section{Forecast discrepancies and adverse outcomes}
For a case--operation unit with $S$ successful attempts among $N$, the empirical frequency is $S/N$. An available model with point prediction $a$ has absolute discrepancy $|a-S/N|$. The tables express this discrepancy in percentage points. A missing forecast has no numerical error. The direct reference receives the same semantic and deployment information, including the corrected Social Network conditions, and is retained even where graph acquisition is unsupported.

Table~\ref{tab:supp-error-summary} groups distinct case--operation units by application and SDK mode, across all treatments and repetitions. Table~\ref{tab:supp-all-discrepancies} identifies all 21 nonzero direct-reference discrepancies. Eighteen of these units also have an available graph-based forecast, whose prediction equals the direct reference; three units have no available graph-based forecast. Shared source/view rows are not counted as new outcomes. The accompanying CSV files retain the unrounded predictions, external counts, state-specific values, unavailable statuses and secondary-control records from which the displayed summaries were formed.

\begingroup\small
\begin{longtable}{@{}L{30mm}L{15mm}R{14mm}R{20mm}R{18mm}R{18mm}R{28mm}@{}}
\caption{Primary support and forecast discrepancies, counted by distinct case--operation unit. Units: all declared units; supported: units with at least one available graph-based forecast; graph misses: supported units with a nonzero discrepancy; direct misses: units with a nonzero direct-reference discrepancy. The maximum is the direct-reference discrepancy in percentage points (pp), including units without graph-based forecasts. The denominator includes all four treatments and three repetitions.}\label{tab:supp-error-summary}\\
\toprule Application & SDK & Units & Supported & Graph misses & Direct misses & Maximum direct discrepancy (pp)\\\midrule
\endfirsthead
\multicolumn{7}{@{}l}{\textit{Table~\thetable{} continued}}\\[3pt]
\toprule Application & SDK & Units & Supported & Graph misses & Direct misses & Maximum direct discrepancy (pp)\\\midrule
\endhead
\midrule\multicolumn{7}{r@{}}{\textit{Continued on the next page}}\\
\endfoot
\bottomrule
\endlastfoot
Petclinic & absent & 48 & 12 & 0 & 0 & 0 \\
Petclinic & partial & 48 & 30 & 0 & 0 & 0 \\
Petclinic & full & 48 & 48 & 0 & 0 & 0 \\
OTel Demo & absent & 36 & 9 & 4 & 5 & 4.21875 \\
OTel Demo & partial & 36 & 9 & 1 & 1 & 0.03125 \\
OTel Demo & full & 36 & 36 & 10 & 10 & 5.59375 \\
Social Network & absent & 36 & 0 & 0 & 0 & 0 \\
Social Network & partial & 36 & 12 & 0 & 2 & 0.0625 \\
Social Network & full & 36 & 36 & 3 & 3 & 0.03125 \\
\end{longtable}

\endgroup

\begingroup\small
\begin{longtable}{@{}L{26mm}L{31mm}L{25mm}R{22mm}R{13mm}R{18mm}C{8mm}@{}}
\caption{All 21 nonzero primary direct-reference discrepancies. A case cell gives SDK mode/repetition and runtime treatment. Successes are successful/total operation attempts, independent of the number of method/view rows. $a_{\mathrm{dir}}$ is the direct conditional prediction; G indicates whether at least one graph-based forecast was available. Every available graph-based forecast equals $a_{\mathrm{dir}}$. Error is absolute discrepancy in percentage points. Operation names refer to \texttt{compose\_post} (Compose), \texttt{browse\_product} (Browse), \texttt{add\_to\_cart} (Cart) and \texttt{checkout} (Checkout). No primary outcomes in these units are unknown or invalid.}\label{tab:supp-all-discrepancies}\\
\toprule Application & Case & Operation & Successes & $a_{\mathrm{dir}}$ & Error (pp) & G\\\midrule
\endfirsthead
\multicolumn{7}{@{}l}{\textit{Table~\thetable{} continued}}\\[3pt]
\toprule Application & Case & Operation & Successes & $a_{\mathrm{dir}}$ & Error (pp) & G\\\midrule
\endhead
\midrule\multicolumn{7}{r@{}}{\textit{Continued on the next page}}\\
\endfoot
\bottomrule
\endlastfoot
Social Network & full/9, NONE & Compose & 2799/3200 & 0.875 & 0.03125 & yes \\
Social Network & full/11, Coroot & Compose & 2799/3200 & 0.875 & 0.03125 & yes \\
Social Network & full/11, DeepFlow & Compose & 2799/3200 & 0.875 & 0.03125 & yes \\
Social Network & partial/9, NONE & Compose & 2798/3200 & 0.875 & 0.0625 & no \\
Social Network & partial/11, NONE & Compose & 2799/3200 & 0.875 & 0.03125 & no \\
OTel Demo & absent/9, Coroot & Browse & 2799/3200 & 0.875 & 0.03125 & no \\
OTel Demo & absent/9, OBI & Cart & 2701/3200 & 0.875 & 3.09375 & yes \\
OTel Demo & absent/9, OBI & Browse & 2702/3200 & 0.875 & 3.0625 & yes \\
OTel Demo & absent/9, OBI & Checkout & 2665/3200 & 0.875 & 4.21875 & yes \\
OTel Demo & absent/11, OBI & Browse & 2799/3200 & 0.875 & 0.03125 & yes \\
OTel Demo & full/10, Coroot & Cart & 2727/3200 & 0.875 & 2.28125 & yes \\
OTel Demo & full/10, Coroot & Browse & 2717/3200 & 0.875 & 2.59375 & yes \\
OTel Demo & full/10, Coroot & Checkout & 2622/3200 & 0.875 & 5.5625 & yes \\
OTel Demo & full/10, DeepFlow & Cart & 2727/3200 & 0.875 & 2.28125 & yes \\
OTel Demo & full/10, DeepFlow & Browse & 2715/3200 & 0.875 & 2.65625 & yes \\
OTel Demo & full/10, DeepFlow & Checkout & 2621/3200 & 0.875 & 5.59375 & yes \\
OTel Demo & full/10, OBI & Checkout & 2793/3200 & 0.875 & 0.21875 & yes \\
OTel Demo & full/11, DeepFlow & Cart & 2797/3200 & 0.875 & 0.09375 & yes \\
OTel Demo & full/11, DeepFlow & Browse & 2795/3200 & 0.875 & 0.15625 & yes \\
OTel Demo & full/11, DeepFlow & Checkout & 2795/3200 & 0.875 & 0.15625 & yes \\
OTel Demo & partial/9, OBI & Checkout & 2799/3200 & 0.875 & 0.03125 & yes \\
\end{longtable}

\endgroup

Agreement with the direct conditional calculation does not establish a forecasting advantage. Disagreement with external outcomes tests the adequacy of the abstraction and remains a reported result. In particular, the cause of the larger retained OTel deviations is unresolved; outcome agreement or disagreement alone does not identify a collector effect. The context-union baseline operates on observed graph evidence and does not demonstrate recovery of per-request causality.

The mixture and state-specific discrepancies are distinct. At state level, 23 case--operation--state units have nonzero discrepancy: 19 have an available graph-based forecast and four have only the direct reference. Of the 19 graph-supported state discrepancies, 14 occur in state $10$, one in state $01$, and four in state $11$. Of the four direct-only state discrepancies, one occurs in state $10$ and three in state $01$. The worked checkout example has 665 successes among 800 attempts in state $10$, a discrepancy of 16.875 percentage points from its prediction of one. Its registered weight of $1/4$ contributes 4.21875 percentage points to the mixture discrepancy. Table~\ref{tab:supp-state-discrepancies} gives the retained state-level extremes and their corresponding mixtures.

\begingroup
\let\supplementoriginallabel\label
\renewcommand{\label}[1]{%
  \def\supplementcandidatelabel{#1}%
  \def\supplementduplicatelabel{tab:state-discrepancies}%
  \ifx\supplementcandidatelabel\supplementduplicatelabel
    \supplementoriginallabel{tab:supp-state-discrepancies}%
  \else
    \supplementoriginallabel{#1}%
  \fi}

\endgroup

\section{Provenance and immutable analytical inputs}
The original case manifest records actual source commits, producing workflows/jobs/attempts, closed inputs and forecasts. The report reader preserved the selected original scientific values without rediscovery or refitting. The corrected Social Network report uses the same original graphs with separately identified corrected forecasts and fresh target observations. A publication source revision is not automatically the source that captured observations or constructed a model.

\begingroup\small
\begin{longtable}{@{}L{36mm}L{120mm}@{}}
\caption{Analytical input and source identities. SHA-256 values identify file contents.}\label{tab:supp-provenance}\\
\toprule Record & Identity\\\midrule
\endfirsthead
\toprule Record & Identity\\\midrule
\endhead
Original fixed-case manifest & \path{docs/v6-full108-completed-result-manifest.json}\par
\hash{23b73280cc0f6386bffe35a29b85fb903fddbdf905a20f61a6643944cdc691a6}\\
Original analytical result & \path{results/literal-results.json}\par
\hash{1153f70032bc36bbc1936e666c8338e0935c24d2c8553b64836e711c9005bf36}\\
Original report producer & Source \hash{c4f066d7c067d5661c3d942aacadb25d3103820d}\par
Workflow run \path{42520421400}\\
Original cost ledger & \path{docs/v6-full108-completed-cost-ledger.json}\par
\hash{198cd5825bea33567353afb5166d552c9a2df963b8f4a86628122de82a4cbf3e}\\
Corrected forecast producer & Source \hash{819e67a8547ca1f3775b328b5a82876e3b39850a}\par
Workflow run \path{42685277599}\\
Corrected forecast plan & \hash{44cd3d7c7fa5e6453564ba1c5cbc838f325f395c5011dff37962389fd9c59c8e}\\
Corrected contract source & \hash{17618a170b0500903a825114dde0a8510530cd09a709786e9eb02081d2311980}\\
All-nine-block barrier & \path{all-nine-blocks-frozen.json}\par
\hash{0e2f69626f9b324c50f323b072ba1faece7a7ffed37b9ab365bacf9b607193ae}\\
Corrected target producer & Source \hash{a53b45d510584b79e6a558945740a17a46bb0830}\par
Workflow run \path{42769442057}\\
Corrected analytical summary & \path{prospective-results-attempt1.json}\par
\hash{d772bd51c199397f5b5d2aaa623d255946cbaf6541b7955ab74a145eaac2a345}\\
Correction publication report & \path{results/correction-publication.json}\par
\hash{d69134c8bf4bc617d3ab5fbe8146617277c7b62cd425801a9c9a8b3928b57f44}\\
Correction report producer & Source \hash{cf41e24567711d982d2a57a6b638d0cb4b4146f9}\par
Workflow run \path{42780232966}\\
Compact source snapshot & \hash{22bac95b046346e5a292776489eac3ecbf179451}\\
Publication analysis & Source \hash{dc2b95d09af57636d9fcd6ace3ee4ae8a739f419}\par
Workflow run \path{43001870142}; analysis script SHA-256\par
\hash{10255b138ccd49e575007b358590066759f3b244c80dc262314e1416e86354fe}\\
Publication analytical output & \path{analysis.json}\par
\hash{13bf5a74527cd7a51939a452567c864432b89b8dce91fc5199a73f437dd92e78}\\
\bottomrule
\end{longtable}
\endgroup

\begin{minipage}{\textwidth}
The selected worked example retains the original OTel/absent/9/OBI training and model packages from source \hash{b913a5df7be680c5e6ede001d0f8eb6fb856a4ed}, run \path{42347357851}. Their original ZIP SHA-256 values are \hash{f910fb03380e6e7ad0c60c4f1e14e8fa7f1858dbcffba9eb37710ba3d9990efb} (training) and \hash{19099f3bdc275fc595c6e819ce010f1da1536152feca3069a1e5333f2389f662} (model). The corresponding frozen forecast manifest is \hash{e820f17d4e670e5b0e22ab0c73f362ec7f991e8c3bec45f09c42386f0724f2b3}.
\par
\end{minipage}

The publication analysis reaggregated the saved original and corrected records in the stated Actions run, without collecting new measurements or building new models. The companion CSV files and the displayed tables use its frozen outputs. The compact builder verified byte counts, SHA-256 identities and archive integrity. Its analytical check loaded the extracted package in an isolated interpreter and recomputed the correction statistics from the packaged summary. That original packaging check did not acquire new application data or rebuild the selected model. The later retained-example replay and its separate source-level evidence checks are documented in Section~\ref{sec:supp-replay}. The deposit does not claim complete historical raw-input replay of all 108 cases. Reexecution of the source code and reaggregation of preserved summaries are distinct from reproducing every historical observation.

\section{Recorded costs and finite accounting scope}
The cost records retain preparation and image work, control and diagnostic runs, capture, baseline/model/simulation processing, fresh validation deployments, transfers, failed and repeated attempts, early stops, report construction and publication work. Reused engineering components and earlier preparation remain identified. Duplicate API aliases of one actual job are retained for provenance and are charged once. Nested phase telemetry is not added to its enclosing worker interval.

The original ledger, closing report work and incremental correction ledger have separate finite cuts. The correction records both the first raw-publication failure and the replacement series. Later successful finalization or delivery work outside a cut is not assigned zero cost. Unknown platform intervals and unmeasured human work remain unknown. Worker elapsed seconds are neither CPU-seconds nor billed time, and concurrent intervals do not represent one sequential experiment duration.

\begingroup\small
\begin{longtable}{@{}L{96mm}R{31mm}R{27mm}@{}}
\caption{Finite cost ledgers and retained breakdowns. Values are recorded known worker elapsed seconds; unknown interval rows remain explicit. The subtotal is the precomputed publication-analysis value. Purpose and phase breakdowns are alternative labels within the original ledger and must not be added to its total or to each other. ``Not stated'' preserves an unspecified count.}\label{tab:supp-costs}\\
\toprule Ledger scope or category & Known seconds & Unknown rows\\\midrule
\endfirsthead
\multicolumn{3}{@{}l}{\textit{Table~\thetable{} continued}}\\[3pt]
\toprule Ledger scope or category & Known seconds & Unknown rows\\\midrule
\endhead
\midrule\multicolumn{3}{r@{}}{\textit{Continued on the next page}}\\
\endfoot
\bottomrule
\endlastfoot
Original pre-reader finite cut & 1,339,757 & 60\\
Closing report finite cut & 1,649 & Not stated\\
Incremental correction finite cut & 262,974 & 1\\
\textbf{Known finite subtotal} & \textbf{1,604,380} & Not combined\\
\midrule
\multicolumn{3}{@{}l}{\textit{Original ledger by purpose (overlaps the phase breakdown)}}\\[2pt]
Source implementation CI & 13,751 & 0\\
Acquisition, modelling and validation & 1,087,724 & 40\\
Technical recovery and saved-input inspection & 233,033 & 20\\
Auxiliary context controls & 3,272 & 0\\
Read-only evidence reader & 939 & 0\\
Maintenance, cleanup and reader registration CI & 1,038 & 0\\
\midrule
\multicolumn{3}{@{}l}{\textit{Original ledger by phase (overlaps the purpose breakdown)}}\\[2pt]
Implementation CI, control or discovery fixture & 14,789 & 0\\
Training acquisition and publication & 435,157 & 0\\
Offline model and baselines & 503,879 & 11\\
External validation and publication & 359,566 & 42\\
Saved-input metadata inspection or skipped gate & 2,243 & 7\\
Auxiliary control acquisition & 3,272 & 0\\
Read-only evidence reader & 939 & 0\\
Preparation, restore or provenance gate & 19,912 & 0\\
\end{longtable}

\endgroup

The table preserves the scope of each reported ledger. It does not establish a complete cost comparison against introducing and maintaining SDK instrumentation, or a cost advantage over the equally informed direct calculation.

\section{Reproducing the publication analysis and reading the data}
\label{sec:supp-reproduction}
This appendix contains Online Resource~1. Online Resource~2 (\path{ESM_2.zip}) contains the publication-analysis script, generated CSV and JSON records, table sources, illustrative model records and the verification evidence. It also includes an exact copy of the deposited \path{code-and-results.zip}: 3,165,309 bytes, SHA-256 \hash{47d9c9a07efc45a9d127a9f0a718a6acc90197c94a878eb6b74548e805b641fa}. The archive root is the deposited repository tree. Its five analytical inputs retain the hashes embedded in the publication script. The supplementary resource therefore supplies the code and inputs needed to regenerate the numerical publication tables. The larger worked-example archive in Zenodo is required for replay from the retained observations.

\subsection{Reaggregation of the archived analytical inputs}
Execute the following as a Bash step in an actual GitHub Actions job with Python~3.12. Make \path{ESM_2.zip} available in the job's workspace before the step. The first commands extract the supplementary resource and its included analytical core into fresh \path{publication/} and \path{core/} directories. GitHub supplies the execution environment and run identifiers used by the script. The output directory must be new or empty; the command uses the current run and attempt to distinguish its output.

\begingroup\small
\begin{verbatim}
cd "$GITHUB_WORKSPACE"
python3 -m zipfile -e ESM_2.zip publication
python3 -m zipfile -e publication/code-and-results.zip core

python3 publication/scripts/build-submission-analysis.py \
  --original \
    "$GITHUB_WORKSPACE/core/results/literal-results.json" \
  --corrected \
    "$GITHUB_WORKSPACE/core/data/prospective-results.json" \
  --correction-report \
    "$GITHUB_WORKSPACE/core/results/correction-publication.json" \
  --cost-ledger \
    "$GITHUB_WORKSPACE/core/docs/v6-full108-completed-cost-ledger.json" \
  --provenance \
    "$GITHUB_WORKSPACE/core/results/provenance.json" \
  --zenodo-doi "10.5281/zenodo.24131735" \
  --out "$GITHUB_WORKSPACE/reproduced-$GITHUB_RUN_ID-$GITHUB_RUN_ATTEMPT"
\end{verbatim}
\endgroup

The script uses the Python standard library. It verifies all five input hashes before reaggregation and retains the fixed selection of 72 original Petclinic/OTel and 36 corrected Social Network cases. A hash mismatch is a failed input-identity check; the embedded pins should remain unchanged. Its outputs include the CSV datasets described below, \path{analysis.json}, \path{case-provenance.json}, a generated-output checksum manifest, and the numerical table sources used for the coverage and discrepancy tables. The manuscript and this supplement apply their own typographic formatting to those saved table values.

For the same script and input bytes, the numerical rows reproduce the archived analysis. The creation timestamp and actual run/source metadata in \path{analysis.json} identify the new execution; recorded input filenames also reflect the extracted paths. Those metadata changes alter the JSON file hash and the corresponding output manifest. Numerical comparisons should use the CSV records and the analytical sections of \path{analysis.json}, with the input hashes checked separately. The archived publication run remains identified in the provenance section above.

\subsection{Dataset keys, units and scope}
The columns \path{system}, \path{sdk}, \path{repetition} and \path{collector} identify a case. Operation-level records add \path{operation}; method/view records also add \path{source} and \path{view}. The latter records share measured outcomes and are not additional repetitions. Probability and absolute-error fields use the unit interval; fields ending in \path{_pp} use percentage points. Latency values are milliseconds and cost values are worker elapsed seconds. A blank numerical cell preserves an unavailable, unknown or inapplicable value according to its status fields; it must not be read as zero. Structured arrays or objects in CSV cells use JSON inside ordinary CSV quoting.

\begingroup\small
\setlength{\parskip}{1.5pt}
\renewcommand{\arraystretch}{1.10}
\begin{longtable}{@{}L{66mm}L{90mm}@{}}
\caption{Publication datasets and their analytical units. The complete method/view records retain both views; primary summaries collapse their verified duplicate primary forecasts.}\label{tab:supp-data-dictionary}\\
\toprule Dataset group & Contents and interpretation\\\midrule
\endfirsthead
\multicolumn{2}{@{}l}{\textit{Table~\thetable{} continued}}\\[3pt]
\toprule Dataset group & Contents and interpretation\\\midrule
\endhead
\midrule\multicolumn{2}{r@{}}{\textit{Continued on the next page}}\\
\endfoot
\bottomrule
\endlastfoot
\path{cases.csv}\par
\path{primary-outcomes.csv}\par
\path{primary-state-outcomes.csv}
& Case census and allocation regime; distinct case--operation outcome counts and direct references; the same outcomes separated by replica state. These files supply the primary denominators without multiplying counts by observation method or graph view.\\
\addlinespace
\path{source-view-results.csv}\par
\path{case-source-summary.csv}\par
\path{primary-summary.csv}\par
\path{all-view-summary.csv}\par
\path{direct-summary.csv}\par
\path{operation-summary.csv}
& Forecast availability, predicted values, measured counts, errors and unavailable reasons for each method/view; corresponding case-level and grouped summaries. The direct comparator and distinct operation-level support/error totals remain identifiable. Database fields preserve the view-specific result.\\
\addlinespace
\path{context-comparison.csv}\par
\path{obi-absent-results.csv}\par
\path{all-primary-discrepancies.csv}
& Ordinary combined/context-union comparison for each case--operation unit; the OBI eBPF-only absent-SDK Petclinic/OTel subset; all 21 nonzero primary direct-reference discrepancies, including those without an available graph-based forecast.\\
\addlinespace
\path{database-outcomes.csv}\par
\path{database-summary.csv}\par
\path{asynchronous-controls.csv}
& Independent database-up/down outcome counts and their direct references; grouped database forecast availability and errors; OTel persistence-control records by phase, including pending states and reuse of paused requests after resumption.\\
\addlinespace
\path{classification.csv}\par
\path{classification-summary.csv}
& Saved operation-family coverage and adjusted Rand index on the known subset, with corresponding case/group summaries. The omitted assignment arrays cannot be reconstructed from these aggregate records.\\
\addlinespace
\path{case-latency.csv}\par
\path{latency-summary.csv}
& Saved per-case duration summaries and summaries across cases. Quantiles are not recomputed from individual attempts or pooled into a new request distribution; these records do not identify causal collector overhead.\\
\addlinespace
\path{costs.csv}
& Finite worker-interval ledgers, unknown interval counts, and overlapping purpose/phase categories. Only rows marked as additive ledger categories contribute to the reported known finite subtotal.\\
\end{longtable}
\endgroup

The accompanying \path{analysis.json} records input identities, selected cases, analytical results and interpretation flags. The \path{case-provenance.json} records retain original producer and input lineage; \path{analysis-manifest.json} identifies the generated output bytes. The \path{selected-example/} records and their receipt illustrate the retained model/predicate/forecast chain. Their inclusion does not supply every input needed to execute that example.

For new acquisition, the deposited V6 protocol and \path{docs/REPRODUCING.md} specify \path{series-v6.yml} for paired blocks and \path{series-isolated-v6.yml} for separate workers. New executions must retain the pre-target forecast freeze and source/run identities. The deposit supports summary reaggregation and new execution; historical raw-input replay covers the selected retained case.

\section{Retained-example replay and source-level evidence}
\label{sec:supp-replay}
The retained OTel Demo/absent/OBI/repetition~9 example was replayed on 11 November 2026 in workflow run \path{43063706864}, from verification source \hash{91c4a8334cea968c00da3f62c23b2f192cdc18e6}. The reader downloaded the complete public Zenodo file without authentication and verified its 1,079,616,429 bytes and SHA-256 \hash{589abd24ce48d28b5b0bdb8fd348ed1ba8efe2efb00abc38f720132cc1ae086b}. It checked the nested archives, the original capture closure and frozen model inputs, then executed the original source \hash{b913a5df7be680c5e6ede001d0f8eb6fb856a4ed} with the pinned Procrustes, Bering and Sheaft revisions. It collected no new application observations and changed no operation contract.

\begingroup\small
\begin{longtable}{@{}L{35mm}L{121mm}@{}}
\caption{Replay checks for OTel Demo/absent/OBI/repetition~9, network-scoped view. Online Resource~2 contains \protect\path{evidence/replay-comparison.json} and the acquisition/execution receipts.}\label{tab:supp-replay-status}\\
\toprule
Check & Recorded result\\
\midrule
\endfirsthead
\multicolumn{2}{@{}l}{\textit{Table~\thetable{} continued}}\\[4pt]
\toprule
Check & Recorded result\\
\midrule
\endhead
Public artifact access & Anonymous download from Zenodo record \texttt{24131735} completed; the full 1,079,616,429-byte archive matched its pinned SHA-256.\\
\addlinespace
Ordered wire input & The ordered HTTP body hashes, byte counts, endpoints and response statuses matched. Both retained sequence files contain 67,781~bytes and have the same SHA-256.\\
\addlinespace
Discovered graphs & Both retained graph forms matched after excluding only \path{metadata.discovered_at} and \path{metadata.source_ref}. The receipt preserves the differing values and both file hashes.\\
\addlinespace
Operation models & Model semantics matched for browse product, add to cart and checkout.\\
\addlinespace
Success predicates & The saved and reproduced predicate objects matched for all three operations.\\
\addlinespace
Operation bindings & All three comparisons matched on the documented admission, required-edge, target, endpoint, resolved-call, assumption and database-support fields. The receipt lists the compared fields.\\
\addlinespace
Forecast semantics & Support, state responses and forecast contents matched for all three operations. Graph and analysis file-reference hashes are retained separately from the semantic comparison.\\
\addlinespace
External outcomes & Original success counts and adverse discrepancies were preserved: browse product 2,702/3,200 and 3.0625~p.p.; add to cart 2,701/3,200 and 3.09375~p.p.; checkout 2,665/3,200 and 4.21875~p.p. No new application outcomes were measured by the replay.\\
\bottomrule
\end{longtable}
\endgroup

The replay preserved the ordered Bering request-body hashes and recovered the same graph contents. Comparison excluded only the process timestamp and the local Bering listener address, and compared entity collections by their identifiers. The operation models, predicates, selected contract-resolution fields and forecast semantics matched for \path{browse_product}, \path{add_to_cart} and \path{checkout}. The original independently measured outcomes and nonzero forecast discrepancies remained unchanged; outcome agreement was not required for replay success.

\subsection{Checkout dependency witnesses}
The checkout predicate requires 14 edge records over nine service pairs. Every required identifier was traced through the archived observed model, quality record, snapshot and resolved operation requirements. Six span/receiver witnesses were retained from the first replay. A subsequent read-only check, run \path{43077620158} from source \hash{6f7ff0386df2b9e9f1016467e8a7bf393b67fa6d}, extracted the remaining eight gRPC SERVER-child witnesses after correcting the verification reader's operation-field selection. This check used the original archived PB files and matched all 16 selected child/parent records to the retained receiver stream. It changed no model, predicate or forecast. Each evidence record retains the source member and hash, trace/span identifiers, receiver byte range and record position.

\begingroup\footnotesize
\begin{longtable}{@{}R{5mm}L{20mm}L{24mm}L{67mm}L{16mm}C{8mm}@{}}
\caption{The 14 required synchronous checkout records for OTel/absent/OBI/repetition~9, network-scoped view. Rows map to \protect\path{edge-NN.json} in Online Resource~2, \protect\path{evidence/edge-evidence/}, with verified span witnesses and exact receiver locators. I/F mark the six initial/eight follow-up witnesses; RPC prefix \texttt{oteldemo.} is omitted. Per-request checkout causality is not established.}\label{tab:supp-checkout-evidence}\\
\toprule
No. & From & To & Protocol and operation & Span kind & Pass\\
\midrule
\endfirsthead
\multicolumn{6}{@{}l}{\textit{Table~\thetable{} continued}}\\[4pt]
\toprule
No. & From & To & Protocol and operation & Span kind & Pass\\
\midrule
\endhead
1 & cart & valkey-cart & Redis: \texttt{EXPIRE} & CLIENT & I\\
2 & cart & valkey-cart & Redis: \texttt{HGET} & CLIENT & I\\
3 & cart & valkey-cart & Redis: \texttt{HMSET} & CLIENT & I\\
4 & cart & valkey-cart & Protocol and operation unspecified & SERVER & I\\
\addlinespace
5 & checkout & cart & gRPC: \path{CartService/GetCart} & SERVER & F\\
6 & checkout & currency & gRPC: \path{CurrencyService/Convert} & SERVER & F\\
7 & checkout & payment & gRPC: \path{PaymentService/Charge} & SERVER & F\\
8 & checkout & product-catalog & gRPC: \path{ProductCatalogService/GetProduct} & SERVER & F\\
9 & checkout & shipping & HTTP: \texttt{POST /get-quote} & SERVER & I\\
10 & checkout & shipping & HTTP: \texttt{POST /ship-order} & SERVER & I\\
\addlinespace
11 & frontend & cart & gRPC: \path{CartService/AddItem} & SERVER & F\\
12 & frontend & cart & gRPC: \path{CartService/GetCart} & SERVER & F\\
13 & frontend & checkout & gRPC: \path{CheckoutService/PlaceOrder} & SERVER & F\\
14 & frontend & product-catalog & gRPC: \path{ProductCatalogService/GetProduct} & SERVER & F\\
\bottomrule
\end{longtable}
\endgroup

These witnesses establish recorded dependencies in the closed source-global training prefix. That prefix also contains preparation, background traffic and other operations. The evidence does not assert that all 14 records belong to one particular checkout request trace. The supplied checkout requirements include the product and cart preparation steps as well as the checkout entry step; the observed graph is resolved against those requirements.

Eight records (\path{edge-05.json}--\path{edge-08.json} and \path{edge-11.json}--\path{edge-14.json}) retain \path{followup_unresolved_reason} from the checkpoint before the additional parent--child search. The final \path{witness_status} reports successful verification, with observations in \path{selected_span_witness} and \path{followup_observed_pair}. The retained reason describes only the earlier checkpoint.

\subsection{Saved context-union differences}
The same read-only verification inspected the three OTel Demo/partial/OBI model archives for repetitions 9--11. Both graph views were checked, giving six representations of three original cases. In each representation, the union added 17 SERVER-child edge records copied from the saved joint model. Every added edge had both endpoint services in the original canonical graph. Existing edge records, services and endpoints remained unchanged. The recorded graph difference agreed exactly with the saved union proof.

\begingroup\footnotesize
\begin{longtable}{@{}C{10mm}L{24mm}R{12mm}C{21mm}C{27mm}C{24mm}L{23mm}@{}}
\caption{Six saved context-union representations for OTel/partial/OBI. Each repetition supplies two views. B/A/C denote browse product/add to cart/checkout. Exact native/eBPF source--span pairs remain unknown; full edge identifiers, hashes and derived selections are in Online Resource~2, \protect\path{evidence/context/}.}\label{tab:supp-context-evidence}\\
\toprule
Rep. & View & Added edges & Existing edges & Services and endpoints & Supported operations & Exact source pairs\\
\midrule
\endfirsthead
\multicolumn{7}{@{}l}{\textit{Table~\thetable{} continued}}\\[4pt]
\toprule
Rep. & View & Added edges & Existing edges & Services and endpoints & Supported operations & Exact source pairs\\
\midrule
\endhead
9 & Raw & 17 & Unchanged & Unchanged & B, A, C & Unknown\\
9 & Network scope & 17 & Unchanged & Unchanged & B, A, C & Unknown\\
10 & Raw & 17 & Unchanged & Unchanged & B, A, C & Unknown\\
10 & Network scope & 17 & Unchanged & Unchanged & B, A, C & Unknown\\
11 & Raw & 17 & Unchanged & Unchanged & B, A, C & Unknown\\
11 & Network scope & 17 & Unchanged & Unchanged & B, A, C & Unknown\\
\bottomrule
\end{longtable}
\endgroup

\begingroup\footnotesize
\begin{longtable}{@{}L{25mm}L{25mm}L{79mm}L{18mm}@{}}
\caption{The 17 added SERVER records for OTel/partial/OBI/repetition~9, network-scoped view, with saved \texttt{trace\_parent} basis and both services already in the canonical graph. B/A/C follow Table~\ref{tab:supp-context-evidence}; a dash denotes no requirement in those bindings. RPC prefix \texttt{oteldemo.} is omitted. Identities and evidence: Online Resource~2, \protect\path{evidence/context/context-added-edges.csv} and its receipt.}\label{tab:supp-context-added-records}\\
\toprule
From & To & Protocol and operation & Required by\\
\midrule
\endfirsthead
\multicolumn{4}{@{}l}{\textit{Table~\thetable{} continued}}\\[4pt]
\toprule
From & To & Protocol and operation & Required by\\
\midrule
\endhead
ad & flagd & HTTP: \texttt{\_OTHER} & ---\\
\addlinespace
checkout & cart & gRPC: \path{CartService/EmptyCart} & ---\\
checkout & cart & gRPC: \path{CartService/GetCart} & C\\
checkout & currency & gRPC: \texttt{\_OTHER} & ---\\
checkout & currency & gRPC: \path{CurrencyService/Convert} & C\\
checkout & flagd & \texttt{connect\_rpc}: \texttt{ResolveBoolean} & ---\\
checkout & flagd & \texttt{connect\_rpc}: \texttt{ResolveInt} & ---\\
checkout & payment & gRPC: \texttt{\_OTHER} & ---\\
checkout & payment & gRPC: \path{PaymentService/Charge} & C\\
checkout & product-catalog & gRPC: \path{ProductCatalogService/GetProduct} & C\\
\addlinespace
email & flagd & gRPC: \texttt{ResolveInt} & ---\\
frontend & cart & gRPC: \path{CartService/AddItem} & A, C\\
frontend & cart & gRPC: \path{CartService/GetCart} & A, C\\
frontend & product-catalog & gRPC: \path{ProductCatalogService/GetProduct} & B, A, C\\
\addlinespace
payment & flagd & HTTP: \texttt{\_OTHER} & ---\\
product-catalog & flagd & \texttt{connect\_rpc}: \texttt{ResolveBoolean} & ---\\
recommendation & flagd & HTTP: \texttt{\_OTHER} & ---\\
\bottomrule
\end{longtable}
\endgroup

The quality records label the selected relations by their stored parent-link evidence. This sidecar check does not reconstruct the exact native/eBPF source identities of each underlying parent/child pair. That field remains explicitly unknown. The support gain is therefore documented by the observed added records and the successful frozen operation predicates, without attributing every added link to an independently verified cross-source span pair.

\subsection{Reaggregation comparison and retained verification files}
The retained-example replay reaggregated the five inputs with the unchanged publication script. Comparison run \path{43081053322} (source \hash{9c0a19a6729322b8eb19f7f7d98d3d43b88a5dd7}) verified 27 outputs against run \path{43001870142}: 20 CSV files, four LaTeX tables and \path{case-provenance.json} were byte-identical. Differences in \path{analysis.json} and its manifest concerned only creation time, run/source identifiers, input paths and derived JSON size/hash. The original container's 33 illustrative model JSON files and their receipt were inventoried separately.

Online Resource~2 retains the replay receipt and comparison, all 14 edge-evidence records, context differences and source hashes, and both analytical output archives with their comparison. Verification scripts and workflows identify producing revisions. The initial witness-selection report, context-reader size-limit report and container-layout comparison failure document technical verification attempts, without adding experimental cases. Later verification jobs fall outside the finite cost ledgers in Section~S7; their receipts retain execution intervals and resource records.

\clearpage
\endgroup

\end{document}